\documentclass[prd,aps,superscriptaddress,showpacs,preprint,preprintnumbers,amsmath,amssymb,nofootinbib]{revtex4}
\usepackage{graphicx}
\usepackage{dcolumn}
\usepackage{bm}

\def\mpi2{m_\pi^2}
\def\mK2{m_K^2}

\newcommand{\bea}{\begin{eqnarray}}
\newcommand{\eea}{\end{eqnarray}}
\newcommand{\be}{\begin{equation}}
\newcommand{\ee}{\end{equation}}

\newcommand{\Dslash}{D\kern-1.5ex /}
\newcommand{\pslash}{p\kern-1.5ex /}
\def\lsim{\raise0.3ex\hbox{$<$\kern-0.75em\raise-1.1ex\hbox{$\sim$}}}
\def\gsim{\raise0.3ex\hbox{$>$\kern-0.75em\raise-1.1ex\hbox{$\sim$}}}

\begin{document}

\bibliographystyle{apsrev}


\newcounter{Outline}
\setcounter{Outline}{1}

\newcounter{Intro}
\setcounter{Intro}{1}

\newcounter{Analytic}
\setcounter{Analytic}{1}

\newcounter{Details}
\setcounter{Details}{1}

\newcounter{Results}
\setcounter{Results}{1}

\newcounter{Conclusions}
\setcounter{Conclusions}{1}

\newcounter{Acknowledgments}
\setcounter{Acknowledgments}{1}

\newcounter{Appendix}
\setcounter{Appendix}{1}

\newcounter{Tables}
\setcounter{Tables}{1}

\newcounter{Figures}
\setcounter{Figures}{1}


\draft

\preprint{BNL-HET-05/20, CTP-3622, RBRC-519}

\title{Calculation of the neutron electric dipole moment with two dynamical flavors of domain wall fermions}

\author{F.~Berruto}
\affiliation{High energy theory group, Brookhaven National Laboratory, Upton, NY 11973}

\author{T.~Blum}
\affiliation{RIKEN-BNL Research Center, Brookhaven National Laboratory, Upton, NY 11973}
\affiliation{Physics Department, University of Connecticut, Storrs, CT 06269-3046}

\author{K.~Orginos}
\affiliation{Department of Physics, PO Box 8795,
College of William and Mary, Williamsburg, VA 23187-8795}
\affiliation{Jefferson Lab,
MS 12H2,
12000 Jefferson Avenue, Newport News, VA 23606}

\author{A.~Soni}
\affiliation{High energy theory group, Brookhaven National Laboratory, Upton, NY 11973}
                 
\date{\today}

\begin{abstract}
We present a study of the neutron electric dipole moment
($\vec d_N$) within the framework of lattice QCD with two flavors of dynamical light quarks. The dipole moment is sensitive to the topological structure of the gauge fields, and accuracy can only be achieved by using dynamical, or sea quark, calculations. However, the topological charge evolves slowly in these calculations, leading to a relatively large uncertainty in $\vec d_N$. 
It is shown, using quenched configurations, that a better sampling of the charge distribution reduces this problem, but because the
CP even part of the fermion determinant is absent, both the topological charge distribution and $\vec d_N$ are pathological in the chiral limit. We discuss the
statistical and systematic uncertainties arising from the topological charge distribution and unphysical size of the quark mass in our calculations and prospects for eliminating them.

 Our calculations employ the RBC collaboration two flavor domain wall fermion and DBW2 gauge action lattices with
inverse lattice spacing $a^{-1}\approx$ 1.7 GeV,  physical volume $V\approx (2$ fm)$^3$, and light quark mass roughly equal to the strange quark mass ($m_{sea}=0.03$ and 0.04). We determine a value of the electric dipole moment that is zero within (statistical) errors,  $|\vec d_N| = -0.04(20)$~e-$\theta$-fm at the smaller sea quark mass. 
Satisfactory results for the magnetic and electric form factors of the proton and neutron are also obtained and presented.

\end{abstract}

\pacs{11.15.Ha, 
	11.30.Er,	
      11.30.Rd, 
      12.38.Aw, 
      12.38.-t  
      12.38.Gc  
}

\maketitle
\newpage


\section{Introduction}
\label{sec:intro}

\ifnum\theIntro=1
%
%

One of the most intriguing aspects of quantum chromodynamics (QCD) is that it allows a gauge invariant interaction term that is separately odd under time-reversal (T) and
parity (P) transformations,
 the so-called $\theta$ term.
The presence of such a term has the profound effect that the
Strong interactions violate the combined symmetry charge-conjugation (C) times P. The existence of P and T violating interactions in the action imply permanent electric dipole moments for fundamental particles. Presently, the most precise search for a permanent electric dipole moment comes from the measurement of the electric dipole moment of the neutron, $\vec d_N$. 
In the Standard Model, the CP-odd phase of the CKM mixing matrix also produces a non-vanishing value for $\vec d_N$, but only beyond
one loop order in the Weak interaction.
Consequently, this contribution to $\vec d_N$ is estimated to
be less than $10^{-30}\, e$-cm, many orders of magnitude
below the current experimental bound\cite{Harris:1999jx},
$d_N = |\vec{d}_N| < 6.3 \times 10^{-26}\,e$-cm
(see also \cite{Bijnens:1996ni}). 
There are recent proposals to improve this bound by two to three orders of magnitude by studying the electric dipole of the deuteron at Brookhaven National Laboratory\cite{Semertzidis:2005} and an isotope of radium ($^{225}$Ra) at Argonne National Laboratory \cite{Ahmad:2005}. The latter is now underway.

Using this experimental bound with theoretical
estimates of $d_N/\theta$ \cite{Baluni:1978rf,Crewther:1979pi,Aoki:1990zz,Pich:1991fq,Pospelov:1999ha,Faccioli:2004jz},
then implies a bound on the value of the fundamental
CP-odd parameter in the QCD action,
$\theta \,\lsim\, 10^{-10}$,
which is deemed to be {unnaturally} small. Since
there is no good reason  for this number to be so
different from unity ({\it i.e.}, a heretofore unknown symmetry in Nature), its minuteness requires
``fine-tuning" of the action.  This is
often termed the {Strong CP problem}. A way around the fine-tuning is the so-called Peccei-Quinn mechanism based on a new (undiscovered) symmetry of Nature which requires that $\theta$ be zero~\cite{Peccei:1977hh,Peccei:1977ur,Wilczek:1977pj,Weinberg:1977ma}.

In this paper we present a calculation of 
$d_N$ in units of $\theta$ within the framework of QCD with two flavors of light quarks using the lattice regularization. 
A preliminary report on this work appears in the proceedings of the Lattice 2005 meeting \cite{Blum:2005} held at Trinity College, Dublin, and 
we note that while finishing this work, a similar study, but in the quenched approximation, has appeared in~\cite{Shintani:2005xg}. 

As explained in Section II, the electric dipole moment is sensitive to the topology of the gauge field, or more specifically, fluctuations of topological charge; thus we focus mainly 
on calculations with dynamical, or sea, quarks. The two flavor 
ensemble of lattice gauge fields that we use was generated by the RIKEN BNL Columbia (RBC) collaboration. Details of these simulations are described in~\cite{Aoki:2004ht}. We find that a precise and accurate calculation requires ensembles with significantly longer evolutions ({\it i.e.}, more independent configurations) than presently available; the topological charge has very long auto-correlations. We expect that longer evolutions will be available in the near future\footnote{The RBC and UKQCD collaborations are jointly beginning extensive simulations with 2+1 flavors of domain wall fermions.}. This situation is to be compared to the quenched, or
zero flavor, case where topological charge can be evolved more efficiently.
The topological charge susceptibility, however, does not vanish as the
valence quark mass approaches the chiral limit, and as we show, neither does the electric dipole moment. This quenched pathology means $d_N$ can only be accurately calculated when the sea quarks are included\cite{Berruto:2004cr}. Not surprisingly, this was found to be true in a recent work using the instanton liquid model\cite{Faccioli:2004jz} where it was argued that the quenched chiral limit of $d_N$ is singular. The partially quenched limit $m_{val}\to 0$, $m_{sea}$ fixed is also singular \cite{O'Connell:2005un}.

Since topology is crucial in the calculation of $d_N$, it may also be important to use lattice fermions that do not spoil certain topological relations to the gauge field with large lattice spacing errors. The axial anomaly in QCD relates the topological charge to the pseudoscalar density; a chiral rotation on the quark fields in the QCD action shifts the CP-odd $\theta$ term between gluon and quark sectors. In order to realize this proper behavior, we use domain wall fermions which are chirally symmetric even when the lattice spacing is non-vanishing.  Thus, this important continuum property of QCD is realized at non-zero lattice spacing,
a feature that is absent for Wilson- and staggered- type fermions.

This paper is organized as follows. Section~\ref{sec:framework} describes the method to calculate $d_N$, Section~\ref{sec:lattice details} gives details of the simulations, in Section~\ref{sec:results} we present our results, and in Section~\ref{sec:conclusions} we summarize the present study and outline future calculations.

\fi


\section{General Analytic Framework}
\label{sec:framework}

\ifnum\theAnalytic=1
%
%

\subsection{Theoretical background}
\label{subsec:theory}

We begin by considering the addition of a {T- and P-odd} (therefore CP-odd) term to the QCD Lagrangian (our conventions are detailed in Appendix A):
\begin{eqnarray}
S_{QCD,\theta}&=& -\theta \int dt \int d^3x \,\frac{g^2}{32\pi^2}\mbox{tr}\left[\epsilon_{\mu\nu\rho\sigma}G^{\rho\sigma}(\vec x,t)G^{\mu\nu}(\vec x,t)\right],
\label{eq:S_theta}
\end{eqnarray}
where $G_{\mu\nu}$ is the gluon field strength and the trace is over (suppressed) color indices. Such a term is allowed by the gauge, Lorentz, and discrete symmetries of QCD.
It is easy to see that this so-called $\theta$ term is odd under P and T transformations since
\begin{eqnarray}
\epsilon_{\mu\nu\rho\sigma}G^{\rho\sigma}G^{\mu\nu}&\sim & \vec E \cdot \vec B.
 \end{eqnarray}
$\theta$ is a fundamental, but unknown parameter of QCD. Remarkably, even though the r.h.s. 
of Equation~\ref{eq:S_theta} can be
written as a total divergence, it does not vanish\cite{Coleman:1978ae} and therefore has physical consequences, most notably
CP violation in QCD. We return to this shortly.

On the other hand, the QCD {Lagrangian} for massless fermions,
\begin{eqnarray}
{\cal L}_{QCD,f} &=& \bar\psi\left(i \Dslash\right)\psi,
\end{eqnarray}
is invariant under chiral transformations of the
quark fields,
\begin{eqnarray}
\psi &\to& (1+i\phi\gamma_5/2)\psi,\\
\bar\psi &\to& \bar\psi(1+i\phi\gamma_5/2),
\end{eqnarray}
but the measure of the path intergral is not\cite{Fujikawa:1979ay},
\begin{eqnarray}
{\cal D}\psi {\cal D}\bar\psi &\to& {\cal D}\psi {\cal D}\bar\psi \,
\exp{\left\{i\,\phi \int d^4x\, 
\frac{g^2}{32 \pi^2}\mbox{tr} \left[\epsilon^{\mu\nu\rho\sigma}G_{\mu\nu}G_{\rho\sigma}\right] \right\}},
\end{eqnarray}
which gives rise to the {Adler-Bell-Jackiw anomaly}. It is well known that this axial anomaly induces observable effects even in the CP even case, the mass of the $\eta^\prime$ and the (relatively) large decay rate for $\pi^0\to \gamma \gamma$ to name just two. In this work we are interested in CP-violating effects, ones that vanish when the $\theta$ term is absent from the Lagrangian.

Choosing $\phi=-\theta$, the $\theta$ term can
be rotated away, or canceled in the action. Recently, Creutz has proposed a scenario in the {\it one-}flavor theory where the $\theta$ term can not be removed, even in the massless limit\cite{Creutz:2004zg,Creutz:2004fi}.
Since we deal with at least two flavors of quarks, we will not consider this possibility further.

If all the quarks are massive, the chiral rotation generates another term in the action that can not be canceled by further field re-definitions,
\begin{eqnarray}
m \bar\psi \psi &\to& m\bar\psi\psi + {i \theta m\bar\psi\gamma_5\psi},
\end{eqnarray}
which is also P- and T- odd. Thus the CP-violating term in the QCD Lagrangian can be moved between the gauge and fermion sectors, but it can not be eliminated. 

Even though it cannot be eliminated, the $\theta$ term can be written as a total derivative, as mentioned above. Still, as is well known, it does not vanish. For QCD
\cite{Coleman:1978ae}
\begin{eqnarray}
\int d^4x \frac{g^2}{32 \pi^2}
\mbox{tr} \left[\epsilon_{\mu\nu\rho\sigma}G^{\mu\nu}G^{\rho\sigma}\right] 
&=& Q,
\end{eqnarray}
where $Q$ is the integral topological charge ($Q=0,\pm1,\pm 2,\dots$) .
Thus the $\theta$ term produces physical effects, like an electric dipole moment for the neutron.

Theoretical calculations naturally yield $d_N$ in units of the unknown fundamental parameter $\theta$. Thus 
to translate the current experimental bound to a constraint
on $\theta$, or to determine $\theta$ should a non-zero value of $d_N$ be found through experiment, requires
evaluation of nucleon matrix elements. The
lattice regularization of QCD provides a first-principles technique for such calculations which we describe after discussing the chiral limit
of $d_N$.

\subsection{Taking the chiral limit}
\label{subsec:chiral}

Consider the QCD partition function for $N_f$ flavors of massive, degenerate, quarks after integrating over the Grassman quark fields,
\begin{eqnarray}
Z &=&
\int {\cal D}{\cal A}_{\mu}\ \mbox{det}[\Dslash(m,\,{\cal A}_{\mu})+i\theta {m}\gamma_5]^{N_f}
\ e^{-S({\cal A}_\mu)},
\end{eqnarray}
where ${\cal A}_{\mu}$ is the gluon field, $\Dslash(m, \,{\cal A}_{\mu})$ is a general covariant Dirac operator for a single quark flavor with mass $m$ as may be found in the continuum or on the lattice.
The choice of degenerate quarks is made for convenience of notation with no loss of generality in the following.
Factoring out $\det\Dslash(m, \,{\cal A}_{\mu})$, the CP-even part of fermion action, and assuming $\theta$ is small,
\begin{eqnarray}
\det{[\Dslash(m, \,{\cal A}_{\mu})+i{ \theta {m}\gamma_5}]} &=&
\mbox{det}[\Dslash(m, \,{\cal A}_{\mu})] \ [1+i{\theta {m}\
\mbox{tr}(\gamma_5 \Dslash
(m)^{-1})}\ ] + {\cal O}(\theta^2).
\label{eq:det approx}
\end{eqnarray}
Next, 
using the spectral decomposition of the inverse Dirac operator,  \begin{eqnarray}
\Dslash(m, \,{\cal A}_{\mu})^{-1} &=& \sum_{\lambda} \frac{|\lambda\rangle\langle\lambda|}{i\lambda+m},
\end{eqnarray}
where $\lambda$ is an eigenvalue and $|\lambda\rangle$ an
eigenvector of $\Dslash$,
and the Atiyah-Singer index theorem, we find
\begin{eqnarray}
\mbox{tr}\left[\gamma_5
\Dslash^{-1}(m_f)\right]&=&\frac{(n_+-n_-)}{m}\\
 &=&\frac{Q}{{m}},
\end{eqnarray}
where $n_\pm$ are the number of right- and left-handed chiral zero-modes of $\Dslash$ for a given gauge field configuration with topological charge $Q$. In the limit ${m}\to 0$, the $\theta$ term does not vanish from the action since the factors of ${m}$ cancel, contradicting our above argument derived in the explicit $m=0$ limit in Section~\ref{subsec:theory}.
Similarly, it is not obvious that the field strength $\theta$ term,
$i \theta Q$, vanishes as $m\to 0$.

The seeming contradiction is easily resolved by examining the role of the usual (CP-even) fermionic action,  
\begin{eqnarray}
\det{\Dslash(m, \,{\cal A}_{\mu})}^{N_f} &=& {\Pi}_j\,  (i\lambda_j + m)^{N_f}.
\end{eqnarray}
As $m\to0$, $Q\neq 0$ configurations are suppressed since they support exact zero-modes of $\Dslash$ with zero eigenvalue. 
In other words, in the chiral limit the $Q\neq0$ configurations represent a set of measure zero, and the distribution of topological charge becomes a delta-function,  $\delta{(Q)}$, with zero width,
$\langle Q^2\rangle = 0$, so the $\theta$ term
effectively vanishes.

The quenched approximation of Equation~\ref{eq:det approx}, $\det{\Dslash(m, \,{\cal A}_{\mu})}=1$, still allows CP-violating physics since
the pseudoscalar density term in the action (or equivalently, the CP odd field-strength term) is not discarded (the same conclusion was reached in \cite{Aoki:1990ix} through a different line of reasoning). However, in light of the arguments just made, the mass dependence of any observable depending on $\theta$ will be completely wrong,
and one should expect significant systematic errors as a result. Indeed, the topological charge susceptibility, $\langle Q^2\rangle/V$, which is closely related to $d_N$ as we have just seen, is well known to be non-vanishing in the pure gauge theory \cite{Witten:1979vv}.
One must also be careful in partially quenched theories ($m_{sea}\neq m_{val}$);
in a recent paper \cite{O'Connell:2005un} it was shown that the
leading valence quark mass 
contribution to $d_N$ in partially quenched chiral perturbation theory is
proportional to $m_{sea} \log m_{val}$. Thus the limit $m_{val}\to 0$, $m_{sea}$ fixed, is singular. 

\subsection{Computational Methodology}

The calculation of the dipole moments centers on the form factors that parameterize the matrix element of the electromagnetic current between nucleon states in the $\theta$ vacuum,
\begin{eqnarray}
\langle p\,^\prime,s^\prime| J^\mu | p,s\rangle_\theta
&=&\bar u_{s'}(p\,^\prime)\Gamma^\mu(q^2)u_s(\vec p)\label{eq:s-matrix element}\\
\Gamma^\mu(q^2) &=&
\gamma^\mu\,F_1(q^2)
\label{eq:form}
\\\nonumber
&&+i\,\sigma^{\mu\nu}q_\nu\,\frac{F_2(q^2)}{2m}\\\nonumber
&&+\left(\gamma^\mu\,\gamma^5\,q^2\,
          - 2m\gamma^5\,q^\mu\right)F_A(q^2)\\\nonumber
&&+\sigma^{\mu\nu}q_\nu\gamma^5\,\frac{F_3(q^2)}{2m},
\end{eqnarray}
where 
\begin{eqnarray}
J^\mu &=& \frac{2}{3}\bar u \gamma^\mu u -\frac{1}{3}\bar d \gamma^\mu d.
\label{eq:em current}
\end{eqnarray}
$q=p'-p$ is the space-like momentum ($q^2<0$) transfered by the external photon, $s$ ($s^\prime$) the spin of the incoming (outgoing) nucleon, $m$ is the nucleon mass, 
$\sigma^{\mu\nu}\equiv i/2\,[\gamma^\mu,\gamma^\nu]$,
and $u_s(\vec p),\,\bar u_s(\vec p)$ are Dirac spinors. 

The four terms on the right-hand side of Equation~\ref{eq:form} are the most general set
consistent with the Lorentz, gauge, and CPT symmetries of QCD.
The insertion of $J^\mu$ probes the electromagnetic structure of the
nucleon; for $q^2\to0$ it is easy to show that
$F_1(0)$ is the electric charge of the nucleon 
in units of $e$ (+1 for the proton, 0 for the neutron),
$F_2(0)$ is the anomalous part of the magnetic moment,
$F_A$ is the anapole moment, and
$F_3(0)$ gives the electric dipole moment.
The last two vanish when $\theta\to 0$.

Later it will be useful to separate $J^\mu$ into its iso-scalar and
iso-vector components.
\begin{eqnarray}
J^\mu &=& \frac{1}{2} J^\mu_V +\frac{1}{6} J^\mu_S\\
J^\mu_V &=& \bar u \gamma^\mu u - \bar d \gamma^\mu d\\
J^\mu_S &=& \bar u \gamma^\mu u + \bar d \gamma^\mu d.
\label{eq:em current components}
\end{eqnarray}

\subsection{Calculating dipole moments on the lattice}

Instead of directly computing matrix elements,
lattice calculations proceed by studying {\it correlation functions} in Euclidean space-time. The desired S-matrix element is obtained
through the usual LSZ reduction formula, but in Euclidean space-time,
usually by relying on the exponential dominance of the ground state, though in principle excited state contributions can also be obtained.
For the case at hand, we study a three-point correlation function where a nucleon with spatial momentum $\vec p$ is created at time $0$ by the interpolating field $\chi^\dagger_N(0,\vec p)$, the current is inserted at time $t$, and then the
nucleon state is annihilated at time $t'$.
\begin{eqnarray}
G^\mu(t^\prime,t)&=&
\langle \chi_N(t',\vec p')\, J^\mu(t,q)\, \chi^\dagger_N(0,\vec p)\rangle.
\end{eqnarray}
Inserting a complete set of relativistically normalized states between each interpolating field and the current and translating all fields to equal times, we obtain
\begin{eqnarray}
G^\mu(t',t)&=&\sum_{s,s'}
\langle 0|\chi_N |p',s'\rangle\langle p',s'|
J^\mu
|p,s\rangle\langle p,s|\chi_N^\dagger|0\rangle
\frac{1}{2E\,2E'}e^{-E'(t'-t)}e^{-E t}+\dots
\label{eq:three-pt-corr-func}\\
&=& G^\mu(q)\times f(t,t',E,E')+\dots,
\label{eq:three-pt-corr-func 2}
\end{eqnarray}
where ``$\dots$" denote excited state contributions which we ignore. 
Note, the correlation function contains the desired S-matrix element,
with no need to analytically continue back to Minkowski space-time.
For convenience, we separate the correlation function into two parts, $G^\mu(q)$ which is a Dirac matrix, and $f(t,t',E,E')$ which collects all the kinematical factors, normalization of states, and time dependence of the correlation function. Color indices have been suppressed.
The interpolating field $\chi_N$ is the conventional one used in most 
lattice simulations, {\it e.g.}, for the proton
\begin{eqnarray}
\chi_P &=& \epsilon_{abc}\left[ u^T_a C \gamma_5 d_b \right] u_c,
\end{eqnarray}
with $a$, $b$, and $c$ color indices.

The states are normalized conventionally,
\begin{eqnarray}
\langle 0 |\chi_N^\dagger|p,s\rangle &=& \sqrt{Z_N} u_s(\vec p),
\end{eqnarray}
and using the spinor relation\footnote{In this section we use Euclidean space conventions for the gamma matrices. See Appendix A for details. $E(\vec p)$ appears without a factor of $i$ since we work in Euclidean time and momentum 3-{space}, that is, our external states are on-shell as they must be.}
\begin{eqnarray}
\sum_{s} u_s(\vec p)\bar u_s(\vec p)&=& E(\vec p)\gamma^t -i\vec\gamma\cdot \vec p+ m
\label{eq:spinor}
\end{eqnarray}
and the projector
\begin{eqnarray}
{\cal P}^{xy}&=&\frac{i}{4}\frac{1+\gamma^t}{2}\gamma^y\gamma^x,
\end{eqnarray}
setting $\vec p\,^\prime=0$ and $\theta =0$,
we find the magnetic form factor $G_M(q^2)$.
\begin{eqnarray}
{\rm tr} {\cal P}^{xy}\,G^{x}(q^2) &=& p_y\,m ( F_1(q^2)+F_2(q^2)) 
\label{eq:magnetic 3pt}\\
{\rm tr} {\cal P}^{xy}\,G^{y}(q^2) &=& -p_x\,m ( F_1(q^2)+F_2(q^2))
\label{eq:magnetic 3pt y} \\
G_M(q^2)&=& F_1(q^2)+F_2(q^2).
\label{eq:magnetic form}
\end{eqnarray}
Similarly,
\begin{eqnarray}
{\rm tr} {\cal P}^{t}\,G^t(q^2) 
&=&  m\,( E+m )\left (F_1(q^2) + \frac{q^2}{(2m)^2} F_2(q^2)\right),
\label{eq:electric 3pt}\\
{\cal P}^{t}&=&\frac{1}{4}\frac{1+\gamma^t}{2},\\
G_E(q^2) &=& F_1(q^2) + \frac{q^2}{(2m)^2} F_2(q^2),
\label{eq:electric form}
\end{eqnarray}
for the electric form factor $G_E(q^2)$.
Throughout this paper we include the
factor $(1+\gamma^t)/2$ in projectors 
to yield the positive parity state (neutron or proton in the CP-even vacuum)
(see, {\it e.g.} \cite{Sasaki:2001nf}).

To determine the desired moment, or form factor,  the factor $f(t,t',E,E')$ appearing in Equation~\ref{eq:three-pt-corr-func 2} must
be removed from the correlation function. This is most easily done
by taking a ratio with another suitably chosen three-point function.
For example, taking the ratio of Equation~\ref{eq:magnetic 3pt} with Equation~\ref{eq:electric 3pt} yields the magnetic dipole moment of the nucleon in the limit $q^2\to 0$.
\begin{eqnarray}
\lim_{t'\gg t\gg0}\,
\frac{1}{p_y}
\frac{{\rm tr}{\cal P}^{xy}G^x_{P,N}(t,t',E,\vec p)}{{\rm tr}{\cal P}^{t}G^t_P(t,t',E,\vec p)}&=&
\frac{1}{p_y}
\frac{{\rm tr}{\cal P}^{xy}G^x_{P,N}(q^2)}{{\rm tr}{\cal P}^{t}G^t_P(q^2)}+\dots
\label{eq:magnetic ratio} \\
&=&
\frac{1}{E+m} \frac{F_1(q^2) + F_{2}(q^2)}{G^{(P)}_E(q^2)}+\dots
\label{eq:gm/ge}\\
\lim_{q\to0}\, \frac{1}{E+m} \frac{F_1(q^2)+F_{2}(q^2)}{G^{(P)}_E(q^2)}
&=& \frac{1}{2m}\times \left\{\begin{matrix}
1+a_{\mu,P}\cr a_{\mu,N}
\end{matrix}\right..
\end{eqnarray}
where we have used $F_1(0) =1$ for the proton and 0 for the neutron,
and $a_\mu=F_2(0)$ is the anomalous part of the moment. $P$ and $N$ denote proton and neutron, respectively, and the denominator is
always evaluated for the proton. 
Because we take ratios corresponding to different components of the electromagnetic current, the finite renormalization constant associated with the local lattice current\footnote{On the lattice only the point-split form of the current is conserved. Here we use a local definition, $\bar\psi\gamma_\mu\psi$.} drops out and need not be calculated.

\subsection{CP violating vacuum, $\theta\neq0$}

In this section we consider the case $\theta\neq0$. First, we must
explain a somewhat subtle issue concerning mixing of the magnetic and electric dipole moment terms in correlation functions arising from the physical mixing of the $\theta=0$ eigenstates in the CP broken vacuum. That is, the neutron mixes with P-odd states when $\theta\neq0$
\cite{Pospelov:1999ha,Aoki:2005b}\footnote{In a preliminary report on this work\cite{Berruto:2004cr}, we did not account for this mixing. We are grateful to M. Pospelov and S. Aoki for pointing this out to us.
The mixing occurs because we work with correlation functions, not directly with matrix elements constructed from CP eigenstates.}. 
This physical mixing of states gives rise to an {\it un}physical mixing of the electric and magnetic dipole moment form factors in correlation functions like those given in Equation~\ref{eq:three-pt-corr-func}.
Generally, the mixing 
can be written as 
a Dirac spinor with phase $e^{i\alpha\gamma_5}$ 
since $\gamma_5 u_s(\vec p) = v_s(\vec p)$, {\it i.e.}, $\gamma_5$ takes a spinor of a given parity into the other.
So, instead of the spinor relation (\ref{eq:spinor}), one obtains
\begin{eqnarray}
\sum_{s,s'} u_{s',\theta}(\vec p) \bar u_{s,\theta}(\vec p) &=&E(\vec p)\gamma_t -i\vec\gamma\cdot \vec p + m e^{2 i \alpha\gamma_5},\label{eq:theta spinor}\\\nonumber
&\approx&E(\vec p)\gamma_t -i\vec\gamma\cdot \vec p + m (1 + 2 i \alpha \gamma_5),\\
u_{s,\theta}(\vec p)&\equiv&e^{i\alpha\gamma_5} u_s(\vec p).
\end{eqnarray}
In the second line we have assumed that $\alpha\ll 1$.
Using $u_{s,\theta}(\vec p)$ in (\ref{eq:three-pt-corr-func}) instead of $u_s(\vec p)$ and proceeding as in the previous section, we obtain
\begin{eqnarray}
{\rm tr} {\cal P}^{xy}\,G^{x}(q^2) &=& 
 p_y\,m ( F_1(q^2)+F_2(q^2)) 
+p_x p_z(\frac{1}{2} F_3 - 2 i m^2 F_A)
 + \frac{\alpha}{2} p_x p_z F_2+ {\cal O}(\theta^2)\\
{\rm tr} {\cal P}^{xy}\,G^{y}(q^2) &=& 
 -p_x\,m ( F_1(q^2)+F_2(q^2)) 
+p_y p_z(\frac{1}{2} F_3 - 2 i m^2 F_A)  + \frac{\alpha}{2} p_y p_z  F_2+ {\cal O}(\theta^2)\\
{\rm tr} {\cal P}^{xy}\,G^{z}(q^2) &=& 
\alpha m ( E-m ) F_1 +\alpha (m ( E-m ) + \frac{p_z^2}{2}) F_2 +
\frac{p_z^2}{2} F_3 + {\cal O}(\theta^2)
\label{eq:f2-f3 three-point function 2}\\
{\rm tr} {\cal P}^{xy}\,G^t(q^2) 
& = & {i p_z} \left( 
\alpha m F_1(q^2) 
+ \alpha \frac{E+3m}{2} F_2(q^2)
+ \frac{E+m}{2} F_3(q^2)\right) + {\cal O}(\theta^2).
\label{eq:f2-f3 three-point function}
\end{eqnarray}
The first two equations are given for completeness; they are not used in our analysis. The last two can
be used to extract the electric dipole form factor $F_3(q^2)$. In particular,
taking the ratio of Equation~\ref{eq:f2-f3 three-point function} with the proton electric form factor correlation function, Equation~\ref{eq:electric 3pt}, we arrive at 
\begin{eqnarray}
\frac{1}{i p_z}\frac{{\rm tr}{\cal P}^{xy}G^t_{N}(t,t',E,\vec p)}{{\rm tr}{\cal P}^{t}G^t_P(t,t',E,\vec p)}
&=& \frac{1}{i p_z}
\frac{{\rm tr}{\cal P}^{xy}G^t_{P,N}(q^2)}{{\rm tr}{\cal P}^{t}G^t_P(q^2)}+\dots\label{eq:f2-f3 ratio}\\
& = & \frac{\alpha m F_1(q^2) 
+ \alpha \frac{E+3m}{2} F_2(q^2)
+ \frac{E+m}{2} F_3(q^2)  }{m(E+m)\, G_E^{(P)}(q^2) }+\dots
\end{eqnarray}
Subtracting the $F_1$ and $F_2$ terms and taking the limit
$q^2\to 0$ yields the electric dipole moment.
\begin{eqnarray}
 \frac{F_3(q^2)}{2 m G_E^{(P)}(q^2)}
&=& \left\{
\frac{1}{i p_z}\frac{{\rm tr}{\cal P}^{xy}G^t_{N}(t,t',E,\vec p)}{{\rm tr}{\cal P}^{t}G^t_P(t,t',E,\vec p)} -
 \frac{\alpha m F_1(q^2) 
+ \alpha \frac{E+3m}{2} F_2(q^2)}{m (E+m) G_E^{(P)}(q^2) }\right\}
\label{eq:electric dipole moment form factor}\\
d_N = \frac{F_3(0)}{2 m} &=& \lim_{q^2\to0} \left\{
\frac{1}{i p_z}\frac{{\rm tr}{\cal P}^{xy}G^t_{N}(t,t',E,\vec p)}{{\rm tr}{\cal P}^{t}G^t_P(t,t',E,\vec p)} -
 \frac{\alpha m F_1(q^2) 
+ \alpha \frac{E+3m}{2} F_2(q^2)}{m (E+m) G_E^{(P)}(q^2) }\right\}.
\label{eq:electric dipole moment}
\end{eqnarray}

The value of the mixing angle $\alpha$ is most easily calculated from the ratio of the zero momentum two-point functions\cite{Aoki:2005b}.
\begin{eqnarray}
\langle\chi_{N^\theta}(t)\chi_{N^\theta}^\dagger(0)\rangle_\theta
&=& \frac{\langle 0|\chi_{N^\theta}|N^\theta\rangle\langle N^\theta|\chi^\dagger_{N^\theta}|0\rangle}{2 m_{N^\theta}} e^{-m_{N^\theta} t}+\dots 
\label{eq:two point}\\\nonumber
&=& Z_{N^\theta}\sum_{s,s'} \frac{u_{s,\theta}(0)\bar u_{s',\theta}(0)}{2 m_{N^\theta}} e^{-m_{N^\theta} t}+\dots,
\end{eqnarray}
where, as usual, ``$\dots$" denotes excited state contributions. Using the spinor relation~(\ref{eq:theta spinor}) and appropriate projectors,
\begin{eqnarray}
\mbox{tr}\frac{1+\gamma_t}{2\cdot 4}~~ \langle\chi_{N^\theta}(t)\chi_{N^\theta}^\dagger(0)\rangle_\theta
& \approx & \,Z_N \,e^{-m_N\,t},\label{eq:two-point even}\\
\mbox{tr}\frac{1+\gamma_t}{2\cdot 4} \gamma_5~~
\langle\chi_{N^\theta}(t)\chi_{N^\theta}^\dagger(0)\rangle_\theta
& \approx & i\,Z_N \,\alpha \,e^{-m_N\,t},\label{eq:two-point odd}
\end{eqnarray}
to lowest order in $\theta$. Note that $Z_{N^\theta}= Z_N+{\cal O}(\theta^2)$ and $m_{N^\theta}= m_N+{\cal O}(\theta^2)$, and as $\theta$ is very small in Nature, we work only to lowest order. 
Of course, the right-hand side of the first equation is nothing but the usual ground state contribution to the nucleon two-point function computed in the CP even vacuum.

Some final remarks are in order. 
CP symmetry for fermions is conventionally defined assuming a real fermion mass. This is the condition that gives the form factors in  
Equation~\ref{eq:form} their usual interpretations, in particular, that the 
electric dipole moment is related to $F_3$. 
Thus, in order to use Equation~\ref{eq:form}, one must work in the standard basis.  On the other hand, following~\cite{Pospelov:1999ha}, one may include in the action the $i\theta' m\bar\psi\gamma_5 \psi$ mass term, arising from a chiral rotation on the quark fields through a particular choice of basis,  in addition to the $-i \theta Q$ term used here.   
There it is shown that $d_N$ depends only on the combination of (renormalized) parameters 
$\bar \theta \equiv \theta+\theta'$, and mixing effects like those described above will differ in just the right way to ensure this is so. In other words, the chiral rotation affects the quark fields in the correlation function as well as in the action. It is only the relative strength, or difference (note the opposite signs of the two terms), of the two contributions that leads to physical effects.  That the physical value of the CP violating parameter must be $\bar\theta$ is clear since through renormalization QCD with bare $\theta \neq 0$ will generate a $\theta^\prime$ term, even if the bare value of $\theta^\prime$ is set to zero in the action (and visa-versa).
Thus, transforming back to the standard basis, one arrives at the combination of 
renormalized parameters $\theta+\theta^\prime$ which couples to $\tilde G G$. 
In the rest of the paper we will not use the notation $\bar\theta$, though it should be understood that when referring to the parameter $\theta$ this is what is meant. 

\subsection{Computing with $\theta\neq0$}

The $\theta\neq0$ action, being complex, is difficult to simulate with conventional lattice methods. However, this problem can be avoided by working in the small $\theta$ limit, 
\begin{eqnarray}
\langle {\cal O}\rangle_\theta &=& \frac{1}{Z(\theta)}
\int {\cal D} {\cal A}_\mu {\cal D}\bar\psi{\cal D}\psi {\cal O}\, e^{-S({\cal A}_\mu) - { i\theta \int d^4x\ \frac{g^2}{32\pi^2}\mbox{tr}\left[G(x)
\tilde{G}(x)\right]}}\\
 &\approx& \frac{1}{Z(0)} 
 \int {\cal D} {\cal A}_\mu {\cal D}\bar\psi{\cal D}\psi  
 (1-i\theta Q) {\cal O}\, e^{-S({\cal A}_\mu) }\\\nonumber
 &=& \langle {\cal O}\rangle -i\theta \langle Q{\cal O}\rangle
 \end{eqnarray}
 where $\cal O$ is a generic operator functional of the fields. Note
$\langle {\cal O}\rangle_\theta$ becomes an expectation value in the CP-even vacuum, the CP-odd part weighted over topological sectors\footnote{In \cite{Guadagnoli:2002nm} it has been proposed to use the pseudo-scalar density as a weight instead. For chirally symmetric lattice fermions that have an index, this is equivalent to
weighting with $Q$. If chiral symmetry is broken, then the two methods should agree in the limit $a\to0$.} 
\begin{eqnarray}
\langle Q{\cal O}\rangle &=&
\sum_{\nu} P(Q_\nu)~ Q_\nu ~ \langle {\cal O} \rangle_{\nu},
\label{eq:Q weighted}
\end{eqnarray}
where $P(Q)$ is the probability that the gauge field configuration has charge $Q$.
As before, the electric dipole moment, or any CP-odd observable, is seen to be closely related to the topological charge, and we expect that any such observable should vanish as $\langle Q^2\rangle /V\to 0$.
In \cite{Diakonov:1995qy} this was shown explicitly for the large $N$ limit.

Chiral perturbation theory shows that $d_N\sim m_\pi^2\log{m_\pi^2}$
\cite{Crewther:1979pi} and $\langle Q^2\rangle /V\sim m_\pi^2$ \cite{Billeter:2004wx}, so each vanishes in the chiral limit, as expected. We will need the formula for the susceptibility,
\begin{eqnarray}
\frac{\langle Q^2\rangle}{V} &=& \frac{f^2 m_\pi^2}{8},
\label{eq:chiral pt chi}
\end{eqnarray}
later to compare to the lattice results. $f\approx f_\pi=130.7$ MeV is the pion decay constant to lowest order in chiral perturbation theory.
 
Finally, the mixing angle $\alpha$ must also vanish as $m_\pi^2 \to 0$ since it is proportional to $\theta$; this will happen as {$\langle Q^2 \rangle/V\to 0$}. It bears repeating that in the quenched
case $\langle Q^2 \rangle/V$ is independent of the quark mass,
implying that $d_N$ and $\alpha$ do not vanish in the chiral limit.

\fi

\section{Lattice Details}
\label{sec:lattice details}

\ifnum\theDetails=1

In this pilot study, the dipole moments of the neutron and proton are computed mainly using an ensemble of $N_f=2$ flavor QCD gluon configurations generated by the RBC collaboration using domain wall fermions (DWF) and the doubly blocked Wilson (DBW2) gauge action. See~\cite{Aoki:2004ht} for details and other basic physics results. Configurations were 
generated for three values of the sea quark mass, $m_{sea}=0.02$, 0.03, and 0.04. In physical units, this range corresponds roughly to $m_s/2 \,\lsim\, m_{sea}\, \lsim \,m_s$, where $m_s$ is the strange quark mass at scale $\mu\approx2$ GeV. This study has focused on the $m_{sea}=0.03$ and 0.04 ensembles, and we obtain results only for equal valence and sea quark masses, $m_f=m_{val}=m_{sea}$.

Two- and three-point nucleon correlation functions have been computed on 440(460) lattices with sea quark mass $m_{f}=0.03$ (0.04), lattice volume $16^3\times 32$ sites, bare gauge coupling $\beta=0.80$ (inverse lattice spacing $a^{-1}\approx 1.7$ GeV), fifth dimension size $L_s=12$, and domain wall height $M_5=1.8$. The lattices were generated with the exact $\phi$ algorithm.
Observables were computed on every 20th trajectory except
for the first 1650 trajectories at $m_{sea}=0.04$ where the separation was 25 lattices.

Besides the dynamical calculations, we have computed two-point functions on 297 quenched DBW2 $\beta=0.87$ ($a^{-1}\approx 1.3$ GeV) configurations,
also generated and studied by the RBC collaboration~\cite{Aoki:2002vt}.
Here, $m_{val}=0.05$ (roughly $m_s$).
 
Throughout, periodic in space and anti-periodic in time boundary conditions are applied to the fermion fields. The gauge fields were generated with periodic boundary conditions.

For the three-point functions, we use the sequential source method described in \cite{Sasaki:2003jh}, and the source and sink are both Gaussian smeared,
\begin{eqnarray}
\chi(\vec x) &=& \left( 1 + \frac{\omega}{4N} \vec\nabla \right)^N\chi(\vec x).
\end{eqnarray}
$N$ is the number of times the smearing kernel acts on $\chi$, and $\omega$ is the width of the Gaussian that results in the limit $N\to \infty$. We took 
$N=30$ and $\omega=4.35$, optimal parameters for quenched Wilson fermions and Wilson gluons at $a^{-1}\approx 2$ GeV \cite{Dolgov:2002zm}. As it turned out,
these were not optimal in the $N_f=2$ DWF case, though they yield satisfactory results when compared to simple wall or point source interpolating fields. On the other hand, these parameters worked exceptionally well for quenched DBW2 $\beta = 0.87$.  
The three-point correlation functions were computed for two source times, alternating on successive lattices, $t=0$ and 15 with sink times $t=10$ and 25, respectively. The electromagnetic current was then inserted in the correlation function between these fixed source and sink times. Correlation functions on successive lattices were 
blocked together to reduce auto-correlations, leading to
220(240) pseudo-independent measurements in each case for 
$m_{sea}=0.03$ (0.04). 
This is roughly the same blocking factor used in \cite{Aoki:2004ht}.

To save computer time, the three-point functions were calculated from the non-relativistic components of the Dirac spinor which is equivalent to using the projector $(1+\gamma^t)/2$ on the source. To calculate the CP odd piece of the two-point function, the full four component Dirac spinor is required since using only the non-relativistic components gives identically zero for the projected two-point function. 

Since the magnetic and electric dipole moment terms in Equation~\ref{eq:form} are proportional to $q^\nu$, to compute the dipole moments, the correlation functions must be calculated for $q\neq0$ and then extrapolated to $q=0$. Attempts to calculate with $q^2=0$, for example by taking a derivative with respect to $q$, do not work on the lattice\cite{Wilcox:2002zt}. For simplicity, we take the out-going nucleon to be at rest, and the incoming nucleon to have spatial momenta $\vec p = -\vec q$. Because high momentum states are more noisy, we restrict ourselves to total momentum $|\vec p\,|^2$ with $p_i = 2\pi (n_i)/L_i $ and $\sum_i n_i^2 \le 4$. Momentum conservation is enforced by summing over the location of the center of the smeared sink and Fourier transforming with respect to the current insertion point. 
\fi


\section{Results}
\label{sec:results}

\ifnum\theResults=1
We begin this section by investigating the topological charge on the
ensemble of $N_f=2$ gauge configurations. Figure~\ref{fig:top charge}  shows the simulation time history of $Q$; evidently there are long autocorrelations, a fact already noted in~\cite{Aoki:2004ht}.
The lower panel corresponds to a quenched simulation ($a^{-1}\approx 1.3$ GeV) where $Q$ fluctuates rapidly. Note the
abscissa is different in the quenched case. The difference 
in fluctuations reflects the
fact that the quenched lattices are separated by 1000 sweeps, whereas the dynamical ones are separated by only five trajectories, owing to the significantly higher cost of the latter.  
In the former case, one sweep consists of one heat-bath plus four over-relaxed hits on each link of the lattice. In the latter, one trajectory = 50 steps of hybrid molecular dynamics evolution of each link plus one global Metropolis accept/reject step. 
We also emphasize that the suppression of tunneling between topological sectors is an algorithmic, not physics, problem which is much worse in the dynamical case due to the smooth hamiltonian evolution of the Monte-Carlo algorithm (see also \cite{Alles:1996vn,Alles:1998jq} for earlier studies of this problem using staggered fermions). The method used to calculate $Q$ uses APE smearing with coefficient 0.45 for twenty steps and an improved definition of the lattice field strength (see \cite{Aoki:2004ht} for details). An even better approach may be to use the overlap definition of the topological charge \cite{Giusti:2001xh,Narayanan:1994gw}, though the precise definition of $Q$ is
probably not the limiting factor.

In Figure~\ref{fig:chi}, the topological susceptiblity $\chi$ is shown for both quenched and $N_f=2$ cases, the former being plotted as a horizontal line since it does not depend on any sea quark mass (the $N_f=2$ results were determined from the data in~\cite{Aoki:2004ht}, the quenched from~\cite{Aoki:2002vt}). $\chi$ and $m_\pi$ are plotted in units of the Sommer scale, $r_0$, to the appropriate power to make each dimensionless. The values for $r_0$ were taken from~\cite{Aoki:2004ht} ($N_f=2$) and~\cite{Aoki:2002vt} (quenched). The interesting feature to note is the significant decrease of the $N_f=2$ value relative to the quenched one. While there may be some sea quark  mass dependence, $\chi$ levels off between $m_{sea}=0.03$ and 0.04. In addition, the statistical errors shown in the figure were estimated by blocking the data in groups of 50 trajectories (10 lattices) and treating the blocks as independent while Figure~\ref{fig:top charge} indicates 
the topological charge has autocorrelations on longer scales. Also shown in Figure~\ref{fig:chi} is the prediction from lowest order chiral perturbation theory (Equation~\ref{eq:chiral pt chi}). It is comforting that this lowest order prediction is consistent with the $N_f=2$ lattice calculation, but because of the caveats just mentioned, the agreement is not yet significant.
Given the close relation between $\chi$ and the quark mass dependence of the electric dipole moment, it does not appear promising that the mass dependence of $d_N$ can be accurately determined from these ensembles; (much) longer evolutions are required. Nevertheless, the 
$m_f=0.03$ and 0.04 calculations may give a relatively good estimate of the magnitude of $d_N$ in QCD where the lightest quark mass is about $m_s$. From Figure~\ref{fig:chi}, this is almost surely {\it not} true for the quenched case.

Next we discuss the CP even and odd parts of the two-point function (Equations~\ref{eq:two-point even}~and~\ref{eq:two-point odd}). Again, working to lowest order in $\theta$ by weighting expectation values with $i Q $ in the latter case, the masses and $Z$ factors obtained from each {\it must} be equal. To reduce statistical errors, we average the forward and backward in time parts of the nucleon propagator and over source time slices 0 and 15 for the $m_{sea}=0.03$ calculation\footnote{We omitted the negative parity states from our earlier discussion and Equation~\ref{eq:two point} for clarity. The backward propagating, negative parity, anti-particle state appears because of the anti-periodic boundary condition in time (see \cite{Sasaki:2001nf}).}. For the usual $\theta=0$ propagator, the former means averaging positive and negative parity states (particle and anti-particle). For $\theta\neq0$, the particle and anti-particle states have the same CP-odd part containing both parities ({\it c.f.}, Equation~\ref{eq:theta spinor}).
Thus, we fit to
\begin{eqnarray}
G_{\rm even}(t,\vec p) &=& A e^{-E(\vec p)\,t}
\label{eq:even 2pt}\\
G_{\rm odd}(t,\vec p) &=& A \left( e^{-E(\vec p)\,t}-e^{-E(\vec p)(N_t - t)}\right),
\label{eq:odd 2pt}
\end{eqnarray}
for the former and latter, respectively, in the range $7 \le t  \le 12$ to avoid excited state contamination.
Ignoring the excited state contributions is justified by the acceptable $\chi^2$/dof of the single particle fits (Tables~\ref{tab:masses}~and~\ref{tab:quenched masses}). For the CP odd case, the $\chi^2$/dof is a bit large in some cases (Tables~\ref{tab:masses theta neq 0}~and~\ref{tab:quenched masses theta neq 0}), but likely for different reasons that are explained below. 
The average over forward and backward propagating states is equivalent to performing a time-reversal transformation on the correlation function which, in turn, is equivalent to averaging over time-reversed gluon configurations. This last step flips the sign of the topological charge on the underlying gluon configuration (recall that the $\theta$ term is odd under time-reversal).
Thus, performing the average of forward and backward correlation functions has the same effect as exactly symmetrizing the topological charge distribution of the ensemble.
 
Table~\ref{tab:masses}~and~\ref{tab:masses theta neq 0} summarize the fits to the two-point function for the $N_f=2$ case. For $m_{sea}=0.04$ the measured values of $E(p)$ are compared to the continuum relativistic dispersion relation, 
\begin{eqnarray}
E(p) &=& \sqrt{\vec p^2 + m_N^2},
\label{eq:dispersion relation}
\end{eqnarray}
in Figure~\ref{fig:energy} with $p_i = 2\pi n /L_i$ and $\sin(p_i)$
($n=0,\pm 1,\pm2,..., \pm L_i-1$),
the latter being the exact lattice momentum for a free lattice fermion with nearest neighbor action.
The agreement is satisfactory for small $|\vec p|$, indicating lattice artifacts are small in this case. $E(p)$ and $m_N$ in
(\ref{eq:dispersion relation}) are taken from the CP even part of the correlation function. Thus, in the following we simply use $p_i=2 \pi n/L_i$ for the momentum. Since this would lead to large ${\cal O}(a^2)$ errors for large $|\vec p|$, we restrict our analysis to the four non-zero lowest values admitted on our lattice, $\vec n=(1,0,0)$, (1,1,0), (1,1,1), and (2,0,0), and permutations.  Since the larger momentum correlation functions are considerably more noisy anyway and would suffer large ${\cal O}(a^2)$ errors with either choice, this
is not a cause for concern. 
Finally, we note that the values of the nucleon masses given in Table~\ref{tab:masses} differ 
by about two (statistical) standard deviations from those reported in \cite{Aoki:2004ht}. 
Different fit ranges and sources were used
($8\le t \le 16$, wall source in \cite{Aoki:2004ht}), and our statistical errors are about two to three times smaller, reflected by our increased statistics.

In Figure~\ref{fig:alpha m=0.03} we show the fitted nucleon mass,
$m_{sea}=0.03$, versus the minimum time slice used in the fit. Values of $m_N$ for both CP even and odd parts of the two-point function are shown. For $t_{min} > 5$, the masses are constant within statistical errors and agree with
each other, as they should to this order in $\theta$ for the CP odd part of the correlation function. Of course, the
errors on the masses from the CP odd part are significantly larger, due to the topological charge re-weighting procedure.
A naive ratio of the two-point functions which gives the mixing angle $\alpha$ is also shown in Figure~\ref{fig:alpha m=0.03} (lower panel). Again, for $t>5$ a suitable constant plateau is evident from which we extract the average value
$\alpha(0.03)= 0.16(2)$ ($6\le t \le 9)$.

In Figure~\ref{fig:alpha m=0.04} we show the results of a similar analysis, but for $m_{sea}=0.04$. This time,
in the whole range of $t_{min}$ the masses clearly disagree outside statistical errors; the difference in central values is roughly ten percent. The value of $\chi^2$/dof for the CP-odd case is roughly two, while in the CP-even case it is less than one (Tables~\ref{tab:masses}~and~\ref{tab:masses theta neq 0}). A naive ratio of the two-point functions, which would give the mixing angle $\alpha$ if the masses were equal, is also shown in Figure~\ref{fig:alpha m=0.04} (lower panel). Although a plateau appears at small $t_{min}$, the ratio appears to decrease approximately linearly with $t$ in the region where the masses are constant but unequal, as expected.
As mentioned above, for the $m_{sea}=0.03$ case, we were able to average over two source times, whereas
for $m_{sea}=0.04$, we only calculated the full four component two point function needed for the CP odd part
from one source time. This suggests that the 
different behavior between the two cases may be due to an improved overlap of
the nucleon fields with the {\it local} charge in the former case and that even more source times would improve
the CP odd signal.

To further study these effects, we have calculated the same two-point functions on  a quenched ensemble of lattices. 
Unfortunately, it was only after the $m_{sea}=0.04$ and quenched calculations that we realized the importance of
having more than one source for the CP odd part of the correlator.
As mentioned already, the topological charge distribution on this ensemble is expected to be correct in quenched QCD because many more Monte-Carlo updates have been performed between measurements. The masses obtained from fits like those in the dynamical case are shown in Figure~\ref{fig:alpha quenched} and given in Tables~\ref{tab:quenched masses}~and~\ref{tab:quenched masses theta neq 0}. Note that the plateaus are quite good in this case, suggesting the Gaussian smearing parameters are near optimal. The masses agree within statistical errors for $t>5$ and the difference of the central values is less than five percent, so now the naive ratio provides a relatively accurate value of the mixing angle, $\alpha=0.214(32)$, where the error is statistical only, and we have averaged over the range $5\le t \le 10$. All of the fits for both CP-even and odd parts of the correlation function have $\chi^2$/dof $<1$. While the quenched result is clearly an improvement over the two flavor one, Figure~\ref{fig:alpha quenched} suggests that an even more accurate sampling of $Q$ is desirable. In Figure~\ref{fig:energy quenched}, $E(p)$ computed from the CP even part of the correlation function is shown. 

The agreement of the masses between the CP even and odd parts of the correlation function is a simple, but non-trivial, check that the distribution of $Q$ is correct. Summarizing our initial studies, the CP odd signal
may be improved significantly by using more sources for the correlator and by increasing the sampling of global topological charge. The former improves the overlap of the nucleon with the local charge fluctuations. 

One way to try to salvage the $m_{sea}=0.04$ calculation is to fit each correlator separately, extract the coefficient of the ground state exponential from the large time region and then take the ratio of these coefficients to determine $\alpha$. One can take the mass in the CP odd case as a free parameter or fix it to the correct value from the CP even case. 
Though this is more correct than just taking the ratio of correlation functions and picking out the incorrect plateau,
it will still yield a value of $\alpha$ with some significant systematic error. After all, the fitted masses differ by about ten percent. Likewise, we may anticipate the value of $\alpha$ to be incorrect by this amount.
Being a bit more systematic, let us say the correlation function itself has been determined close its actual value. Then we
make a ten percent error in the amplitude (and therefore $\alpha$) since the fitted mass is ten percent too low. Now, a reasonable guess may be that the correlation function is actually determined to roughly ten percent of its correct value. Taking both factors into account, we arrive at a systematic uncertainty in $\alpha$ of about 10-20\%. Following this procedure gives $\alpha = 0.07(2)$ and 
$0.16(2)$ for $m_{sea}=0.04$ and 0.03, respectively
(statistical error only). Here, $m_N$ in the CP odd part of the correlation function has been fixed to the CP even value, and $\alpha$ is from the fit with $t_{min} = 7$. 
The agreement with the simple ratio method in the case $m_{sea}=0.03$ 
is a nice check of both methods.
If $m_N$ is left as a free parameter, the resulting value of $\alpha$ is about 50\% lower (Figure~\ref{fig:alpha-fixed}) for $m_{sea}=0.04$, but the same at 0.03 with larger error,
$\alpha=0.18(4)$. So while there is likely significant uncertainty in the value of $\alpha$ at $m_{sea}=0.04$, the
results for the lighter mass appear stable and satisfactory
\footnote{While finishing this manuscript, we noticed a discrepancy between our
value for $\alpha$ in the quenched case and that reported in
\cite{Shintani:2005xg}. 
Our definition of the mixing angle differs from theirs by a factor of 1/2
(compare Equation~\ref{eq:theta spinor} with Equation 20 in \cite{Shintani:2005xg}). However, the quoted value in \cite{Shintani:2005xg} is numerically equal to ours, within statistical errors.  
Unfortunately, communication with the authors of \cite{Shintani:2005xg} and
further checks of our definitions, code, and analysis of the two point functions has not yet resolved this discrepancy.}.

The ratios of the three-point correlation functions given in Equations~\ref{eq:magnetic ratio} and~\ref{eq:f2-f3 ratio} are shown in Figures~\ref{fig:magnetic ratio}~-~\ref{fig:f2-f3 ratio} for each value of $q^2$ and both sea quark masses. For magnetic form factors we average results using Equations~\ref{eq:magnetic 3pt} and \ref{eq:magnetic 3pt y}.
Figure~\ref{fig:edm ratio} shows the ratio $F_3(q^2)/(2 m G^{(P)}_E(q^2))$ using the value of $\alpha$ determined above. 
Despite the flat plateaus shown in the figures, some excited state contamination may still be present at small and large times, given the fitted masses in Figure~\ref{fig:alpha m=0.04}. For some cases, involving larger momentum transfer, the plateaus show an oscillation, presumably due to insufficient statistics, or possibly excited state contamination.
Averages over time slices $4\le t \le 6$ are also given in Table~\ref{tab:average ratios}. The ratios $F_{1,2}(q^2)/G^{(P)}_{E}(q^2)$ and $F_3(q^2)/(2 m G^{(P)}_E(q^2))$ are summarized in Table~\ref{tab:average F ratios}. The value of the $F_1$ ratio for the neutron approaches zero, as required by electric charge conservation. For the proton, the ratio trivially approaches one
since it goes to $F_1(0)/F_1(0)$ (Equations~\ref{eq:magnetic form} and \ref{eq:electric form}), but at the least serves as a check on our evaluation of the three-point functions. 

As $q^2\to 0$, the $F_2(q^2)$ ratios yield the anomalous magnetic moments of the nucleons.  For each value of $q^2$ the magnitudes for the neutron and proton are equal within errors; this should be true for the iso-vector contributions, assuming iso-spin is not broken which is true in our calculation. 
Evidently the iso-scalar contribution from the connected diagrams is zero, or smaller than our statistical errors;
we have not included the disconnected valence quark loop diagrams in the three-point functions which contribute only to the matrix element of the iso-scalar piece of the electromagnetic current.
The values at the lowest value of $q^2$ are not far off from the well known experimentally measured values $a^{(P)}_\mu=1.79$ and $a^{(N)}_\mu=-1.91$~\cite{Eidelman:2004wy}. To show the momentum dependence of these ratios is mild and to compare to experiment~\cite{Jones:1999rz}, we have also plotted the ratio of the magnetic form factors to the electric form factor of the proton, $G_M(q^2)/G_E^{(P)}(q^2)$, as well as the dipole moments in Figure~\ref{fig:gm/ge}. Note, the quark mass dependence of this ratio is also small, mainly showing up as shift in $q^2$ for larger values of $q^2$. In fact,  it is interesting that the last two $m_{sea}=0.03$ points seem to smoothly fill in the 
large gap between the last two $m_{sea}=0.04$ points. By comparing to experiment, one sees that lattice artifacts are
becoming significant for $q^2\,\gsim \,1.2$ (GeV$^2$). 
The agreement with the experimental form factor ratio for the proton at smaller values of $q^2$ is quite satisfactory, in magnitude and $q^2$ dependence, but may be fortuitous since our calculation does not include electromagnetic effects or disconnected valence quark loop contributions, is done at relatively heavy quark mass, and we have not taken the continuum or infinite volume limits.

Finally, we turn to the CP odd form factor $F_3(q^2)$. The $q^2\to 0$ limit of the $F_3(q^2)$ ratio yields $|d_N|$. 
From the last two columns in Table~\ref{tab:average F ratios}, it can be seen that our value for the form factor is 
consistent with zero within errors for both quark masses and both components of the electromagnet current, except for the second lowest value of $q^2$ for each quark mass, where the central value is roughly two standard deviations from zero ($\gamma^t$ component only). The statistical
errors are much larger for the $z$ component of the electromagnetic current 
(Equation~\ref{eq:f2-f3 three-point function 2}) compared to 
the $t$ component (Equation~\ref{eq:f2-f3 three-point function}), hence we discuss only the latter from now on.
The statistical errors for the lighter quark mass are about twice the size for the heavier.
For the three point correlation functions, we have averaged over two source-sink time slice combinations, with sources separated by 16 time slices, or one-half the lattice in the time direction. Thus the increased error at light quark mass may reflect the (usual) greater fluctuations in the quark propagator as the quark mass is reduced. We are also mindful
that lowering the quark mass suppresses the evolution of the topological charge and charge density and likely  
plays a role as well. Again we emphasize that future calculations may benefit (in terms of smaller statistical errors) from
improving the correlation of the nucleon fields and the local charge density by using many source-sink combinations instead of the two used here. Indeed, the statistical error on the ratio already decreased by about a factor of two by using two sources.

Given the large statistical uncertainties, we simply estimate a bound on $d_N$ 
from the error on the $F_3(q^2)$ ratio evaluated at the lowest value of momentum transfer for $m_{sea}=0.03$.
Though the error on the values at $m_{sea}=0.04$ are a factor of two smaller, we do not use these values because
of the possibly large systematic error in the value of $\alpha$ extracted from the two point correlation function.
We find
\begin{eqnarray}
d_N^{\rm lat} &=& \frac {F_3(0)}{2m} \\
&\approx& \frac{F_3(0.399)}{2m G_E(0.399)} \\
|d_N^{\rm lat}| &<& 0.2.
\end{eqnarray}
The mild $q^2$ dependence for the $F_1$ and $F_2$ ratios leads us to believe that this is not a terrible approximation. Our conventions which are the same as those in \cite{Crewther:1979pi,Aoki:1990zz} have lead to a negative central value of $d_N/\theta$ at $m_{sea}=0.03$ and positive at 0.04. Of course, both are consistent with zero within errors. 
We simply mention this to call attention to our careful treatment of the many sources of convention-based signs in this calculation (see section II and the appendix) and so others may compare their calculations to this and future ones.

$d^{\rm lat}_N$ is given in inverse units of  $a\theta$. In physical units, the above bound reads
\begin{eqnarray}
d_N \,\lsim\, 0.02\,\theta\,e~{\rm fm}.
\end{eqnarray}
This upper bound is roughly an order of magnitude larger than (but not inconsistent with) the central value computed from sum rules~\cite{Pospelov:1999ha} and a factor of four times as large as the pion loop contribution~\cite{Crewther:1979pi}.
Also, the above bound is roughly the same size as the result found
in a recent quenched calculation~\cite{Shintani:2005xg}; however, see $^7$.



\fi


\section{Conclusions}
\label{sec:conclusions}

\ifnum\theConclusions=1
%
%

The dipole moments of the proton and neutron have been calculated using lattice QCD. In particular we have focussed on the electric dipole moment of the neutron. Using two flavor QCD, we obtained a rough bound, $d_N \,\lsim\, 0.02\, e\,\theta\,{\rm fm}$, from the statistical error on the central value which was zero within this error. This bound is somewhat larger than previous model calculations, about the same magnitude as found in a recent quenched calculation~\cite{Shintani:2005xg}, and is only a crude estimate given additional significant systematic uncertainties associated with the topological charge distribution and quark mass and momentum dependence of $F_3(q^2)$.

The ratio of magnetic to proton electric form factors were found to be in good agreement with experiment (Figure~\ref{fig:gm/ge}), given the single lattice spacing, volume, and relatively heavy quark masses used in our calculation. This ratio exhibits only a mild dependence on the momentum transfer $q^2$ and quark mass.

Because $d_N$ arises from the CP-odd term in the action, $\int d^4 x G\tilde G $, it is sensitive to the topological charge distribution. The method outlined here to calculate CP odd observables requires re-weighting correlation functions with
topological charge that would otherwise vanish. This suggests that reducing the statistical errors in such calculations can be
achieved by improved sampling of the topological charge, and also by increasing the number of quark sources so the overlap
between the nucleons and the local topological charge density is enhanced. The former was shown to work in the quenched 
approximation, while evidence for the latter came from the $m_{sea}=0.03$ calculation of the CP odd part of the two point correlation function. 

As discussed in Section~\ref{sec:framework}, in the quenched case $d_N$ does not have the correct (physical) quark mass dependence. In the two flavor case, the quark mass dependence is correct, and $d_N$ vanishes in the chiral limit from the presence of the CP even part of the fermion determinant. Future lattice calculations will approach the chiral limit, which will suppress $d_N^{\rm lat}$ even further. Chiral perturbation theory predicts a leading $m_\pi^2\log m_\pi^2$ term \cite{Crewther:1979pi}, but non-leading $m_\pi^2$ terms may also be important\cite{Aoki:1990zz,Pich:1991fq,Pospelov:1999ha}.
A recent calculation in partially quenched chiral perturbation theory \cite{O'Connell:2005un} may help with the needed extrapolations. It is also interesting to note the prediction in that paper for the leading valence quark mass
dependent term, $m_{sea} \log m_{val}$. Thus the limit $m_{val}\to 0$, $m_{sea}$ fixed, is singular. 

$N_f=2+1$ flavor domain wall fermion 
calculations just begun jointly by the RBC and UKQCD collaborations will attempt to address the two most pressing deficiencies of the present calculation, poor statistics for the topological charge and the quark mass dependence of $d_N$.
We plan to implement the two main lessons learned from this study in the new one, namely longer evolutions of the gauge fields and more quark sources to overlap with the local charge density. Both should improve the statistical errors on CP odd correlation functions. Finally, we will investigate the use of twisted boundary conditions~\cite{deDivitiis:2004kq} to essentially eliminate, or at least 
drastically reduce, the extrapolation of the form factors to $q^2=0$.

\fi


\bibliography{paper}


\ifnum\theAcknowledgments=1
\begin{acknowledgments}

We thank our colleagues in the RBC collaboration, in particular, S. Aoki, and N. Christ, 
for many helpful and interesting discussions.

The numerical computations reported here were done on the 400 Gflops QCDSP
supercomputer at Columbia University and the 600 Gflops QCDSP supercomputer at the RIKEN BNL Research Center. We thank RIKEN, Brookhaven National Laboratory and the U.S. Department of Energy for providing the facilities essential for the completion of this work.

This research was supported by the RIKEN BNL Research
Center (Blum, Orginos) and US Department of Energy (DOE) grants DE-AC02-98CH10886 (Berruto, Soni), DF-FC02-94ER40818 (Orginos),
and DE-AC05-84ER40150 (Orginos). Blum has also received support through a DOE  Outstanding Junior Investigator award, DE-FG02-92ER40716.

\end{acknowledgments}
\fi


\ifnum\theAppendix=1
\appendix
%
%
\newcommand{\imag}{i}

\section{Conventions}
In Minkowski space it is conventional to define the chiral basis
with metric $g^{\mu\nu}$ (signature: 1,-1,-1,-1) as\\

 \begin{tabular}{cc}
 $\gamma^0 = \left(\begin{matrix} 0 & 1\cr 1 &0 \end{matrix}\right)$ & 
 $\gamma^i = \left(\begin{matrix}0 & \sigma^i\cr -\sigma^i &0\end{matrix}\right)$,
 \end{tabular} 
 
 where 
 \begin{tabular}{ccc}
 $\sigma^1 = \left(\begin{matrix}0 & 1\cr 1 &0\end{matrix}\right)$ & 
 $\sigma^2 = \left(\begin{matrix}0 & -i \cr i &0\end{matrix}\right)$ &
 $\sigma^3 = \left(\begin{matrix}1 & 0\cr 0&-1\end{matrix}\right)$
 \end{tabular} are the Pauli matrices.\\

In reality, our code uses slightly different conventions. With the replacement $\vec\gamma_E = -i\vec\gamma$ and $\gamma^4_E=\gamma^0$,
 \begin{eqnarray}
-i\gamma^1  &=&  -\gamma^1_{E}\\
-i\gamma^2  &=& \gamma^2_{E}\\
-i\gamma^3  &=&  -\gamma^3_{E}\\
\gamma^0  &=& \gamma^4_E,
\end{eqnarray}
{\it i.e.}, our definitions of $\gamma^x$ and $\gamma^z$ have the opposite sign of the usual ones (still, 
$\gamma_E^5\equiv \gamma^1_E\gamma^2_E\gamma^3_E\gamma^4_E =-\gamma^5$).
Of course, the results of Eqs.~\ref{eq:magnetic 3pt},~\ref{eq:electric 3pt},~\ref{eq:f2-f3 three-point function}, and \ref{eq:electric dipole moment form factor} do not depend on the choice of metric or signs for the gamma matrices as long as the projectors are also modified accordingly. We have checked this explicitly by comparing results from our conventions to ones using the above conventional chiral basis.
In the above the subscript ``E" stands for Euclidean and is dropped in the main text.

In Minkowski space the gluon action takes the form
\begin{eqnarray}
{\cal L} &=& -\frac{1}{4}G^{\mu\nu}G_{\mu\nu} -\theta \frac{g^2}{16 \pi^2} G^{\mu\nu}\tilde G_{\mu\nu}\\
S &=& \int dt \int d^3x \,{\cal L} \\
&=& -\int dt \int d^3x \left(\frac{1}{4}G^{\mu\nu}G_{\mu\nu} 
+\theta \frac{g^2}{16 \pi^2} G^{\mu\nu}\tilde G_{\mu\nu}\right)
\end{eqnarray}
Continuing to Euclidean space,
\begin{eqnarray}
t &\to &-i\tau\\
G^{0i} &\to& i G^{4i}\\
G_{0i} &\to& -i G^{4i}\\
G^{ij}&\to& -G^{ij}\\
G_{ij} &\to & -G^{ij}.
\end{eqnarray}
The continuation of the field strength term can be worked out
from
\begin{eqnarray}
G^{\mu\nu} &=& \partial^\mu A^\nu - \partial^\nu A^\mu -i g [A^\mu,A^\nu]\\
A^0 &\to& -i A^4\\
\tilde G_{\mu\nu} &=& \epsilon_{\mu\nu\alpha\beta} G^{\alpha\beta},
\end{eqnarray}
and similarly for the (contra)covariant (dual)field strength. $A^\mu$ is the four-vector potential and $\epsilon_{0123}=\epsilon_{1234}\equiv+1$.  Then 
\begin{eqnarray}
\exp{i S} &\to& \exp{\left\{-\int d\tau \int d^3x \left(\frac{1}{4}G^{\mu\nu}G^{\mu\nu} 
+i\theta \frac{g^2}{16 \pi^2} G^{\mu\nu}\tilde G^{\mu\nu}\right)\right\}}\\
&=& \exp\{ -S_E\}.
\end{eqnarray}
Thus, the sign of the $\theta$ term does not change upon analytic continuation to Euclidean space.
\fi


\ifnum\theTables=1
\clearpage
\begin{table}[!htb]
\caption{Masses and energies from fits to the CP even part of the nucleon two-point function ($\theta=0$). Results are averaged over all possible permutations of the lattice momentum given in the first column, including both positive and negative directions. The fit range is $7\le t \le 12$. $m_{sea}=0.03$ (upper) and $m_{sea}=0.04$ (lower).}
\label{tab:masses}
\begin{ruledtabular}
\begin{tabular}{ccccc}
 $\vec p$ & $E(p)$ (error) & $\chi^2/$dof & dof \\\hline
(0,0,0) & 0.8646 (53) & 0.74 &4\\
(1,0,0) &  0.9453 (65) & 1.6 & 4\\
(1,1,0) &  1.031 (10) &  1.3 & 4\\
(1,1,1) &  1.104 (18) &  2.1 & 4 \\
(2,0,0) &  1.140 (29) &  1.3 & 4 \\\hline
(0,0,0) & 0.9264 (54) & 0.31 &4\\
(1,0,0) & 1.0021 (61) & 0.39 &4\\
(1,1,0) & 1.0685 (88) & 0.25 &4\\
(1,1,1) & 1.124 (16) &  0.45 & 4 \\
(2,0,0) & 1.217 (30) & 0.67 &4\\\hline
\end{tabular}
\end{ruledtabular}
\end{table}

\begin{table}[!htb]
\caption{The same as Table~\ref{tab:masses}, except for the CP odd part of the two-point function ($\theta\neq0$). Note, the fit range for
momentum (1,1,1) and $m_{sea}=0.03$ begins at $t=6$.}
\label{tab:masses theta neq 0}
\begin{ruledtabular}
\begin{tabular}{cccc}
 $\vec p$ & $E(p)$ (error) & $\chi^2/$dof & dof \\\hline
 (0,0,0) & 0.881 (34) & 0.50 & 4 \\
(1,0,0) & 0.974 (65) & 0.77 & 4 \\
(1,1,0) & 1.10 (18) &  2.27 & 4 \\
(1,1,1) & 1.17 (42) &  0.91 & 5 \\
\hline
(0,0,0) & 0.814 (36) & 1.85 &4 \\
(1,0,0) & 0.873 (54) & 1.90 &4 \\
(1,1,0) & 0.950 (86) & 2.48 &4 \\
(1,1,1) & 1.18(16) & 0.93 &4\\\hline
\end{tabular}
\end{ruledtabular}
\end{table}

\begin{table}[!htb]
\caption{The same as Table~\ref{tab:masses}, except for the quenched ensemble.}
\label{tab:quenched masses}
\begin{ruledtabular}
\begin{tabular}{cccc}
 $\vec p$ & $E(p)$ (error) & $\chi^2/$dof (dof) \\\hline
(0,0,0) & 1.0217 (44) & 0.46 (4) \\
(1,0,0) & 1.0889 (52) & 0.53 (4) \\
(1,1,0) & 1.1538 (70) & 0.24 (4) \\
(1,1,1) & 1.2152 (110) & 0.14 (4) 
\end{tabular}
\end{ruledtabular}
\end{table}

\begin{table}[!htb]
\caption{The same as Table~\ref{tab:masses}, except for the quenched ensemble and $\theta\neq0$.}
\label{tab:quenched masses theta neq 0}
\begin{ruledtabular}
\begin{tabular}{cccc}
 $\vec p$ & $E(p)$ (error) & $\chi^2/$dof (dof) \\\hline
(0,0,0) & 0.996 (30) & 0.97 (4) \\
(1,0,0) & 1.054 (31) & 0.36 (4) \\
(1,1,0) & 1.123 (47) & 0.71 (4)
\end{tabular}
\end{ruledtabular}
\end{table}

\begin{table}[!htb]
\caption{Ratios of three-point functions given in Equations~\ref{eq:magnetic ratio}~and~\ref{eq:f2-f3 ratio}. In the limit $q^2\to 0$, these yield the dipole moments, except in the case of the electric dipole moment (last column) where the mixing with the $F_1$ and $F_2$ terms has not been subtracted. $m_{sea}=0.03$ (upper) and $m_{sea}=0.04$ (lower).}
\label{tab:average ratios}
\begin{ruledtabular}
\begin{tabular}{cccc}
$q^2$ (GeV$^2$) & proton (magnetic) & neutron (magnetic) & neutron (electric) \\\hline
0.399 & 1.524 (54)  & -0.957 (37) & 0.35 (19)\\
0.824 & 1.624 (65)  & -1.030 (43)& -0.11 (22)\\
1.183 & 1.751 (129)& -1.075 (84) & 0.43 (36)\\
1.363 & 1.408 (176)& -0.952 (131) & 0.30 (53)\\\hline
0.401 & 1.438 (30)  & -0.913 (23)  & -0.045 (87) \\
0.753 & 1.451 (34)  & -0.936 (26) & 0.066 (96) \\
1.044 & 1.594 (68)  & -1.032 (50) &-0.035 (160) \\
1.538 & 1.193 (104) & -0.730 (69) & -0.393 (249) 
\end{tabular}
\end{ruledtabular}
\end{table}

\begin{table}[!htb]
\caption{Form factors normalized by the electric form factor 
of the proton, $G_E(q^2)$ (Equation~\ref{eq:electric form}). In the limit $q^2\to0$, the values in the $F_2$ columns yield the anomalous magnetic moments and in the last two columns, the electric dipole moment of the neutron using currents and projectors defined in 
Equations~\ref{eq:f2-f3 three-point function 2} and~\ref{eq:f2-f3 three-point function}. $m_{sea}=0.03$ (upper) and $m_{sea}=0.04$ (lower). }
\label{tab:average F ratios}
\begin{ruledtabular}
\begin{tabular}{ccccccc}
$q^2$ (GeV$^2$) & \multicolumn{2}{c}{$F_1(q^2)/G_E(q^2)$} & \multicolumn{2}{c}{$F_2(q^2)/G_E(q^2)$} & \multicolumn{2}{c}{$(F_3(q^2)/2\,m)/G_E(q^2)$}  \\
&proton& neutron & proton & neutron & \multicolumn{2}{c}{neutron} \\\hline
0.399  & 1.0784 (63)  & -0.034(10) & 1.680 (98)  & -1.698 (68) &  -0.04 (20) & 0.49 (45) \\
0.824  & 1.183 (15)  & -0.1261 (211) &  1.90 (12) & -1.827 (82) &  0.45 (23) & 1.56 (73)\\
1.183   &  1.297 (40) & -0.219 (42) &  2.15 (23) & -1.90 (15) &  -0.08 (36)  & -0.74 (1.73)\\
1.363   & 1.251 (57)  & -0.176 (71) &  1.57 (31) & -1.73 (25) & 0.02 (53)  & -0.39(73)\\\hline
0.401 & 1.0695 (42) & -0.045 (7) & 1.703 (59) & -1.715 (46) & 0.087 (95) & 0.12(27)\\
0.753 & 1.1349 (89) & -0.081 (13) & 1.760 (66) & -1.785 (52)& 0.20(10) & -0.18(41)\\
1.044 & 1.2181 (223) & -0.102 (27) & 2.050 (129) & -2.013 (101)& 0.12 (16) & -1.29 (75)\\
1.538 &1.2107 (414) & -0.156 (45) & 1.345 (195)& -1.409 (136)  & -0.29 (25) & 0.55 (38)
\end{tabular}
\end{ruledtabular}
\end{table}

\clearpage
\pagebreak

\fi


\ifnum\theFigures=1
%
%
\clearpage

\begin{figure}[b]
\begin{center}
\includegraphics[width=\textwidth]{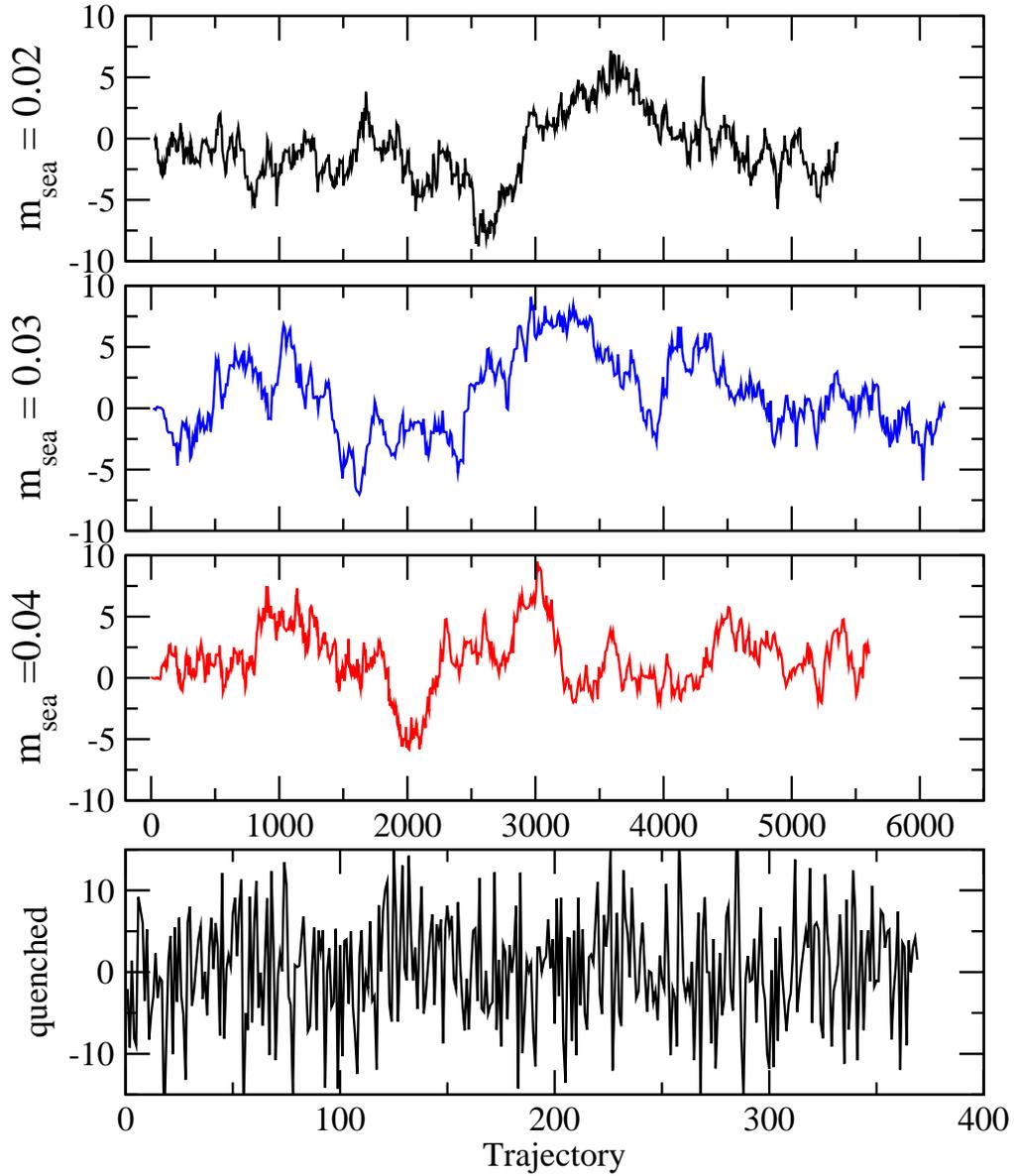}
\end{center}
\caption{Topological charge, $Q$. For the $N_f=2$ simulations, $Q$ for every fifth trajectory is shown, while for the quenched case $Q$ has been measured on lattices separated by 1000 sweeps. The $N_f=2$ plots are reproduced from~\cite{Aoki:2004ht}.}
\label{fig:top charge}
\end{figure}
\clearpage
\newpage

\begin{figure}
\begin{center}
\includegraphics[width=\textwidth]{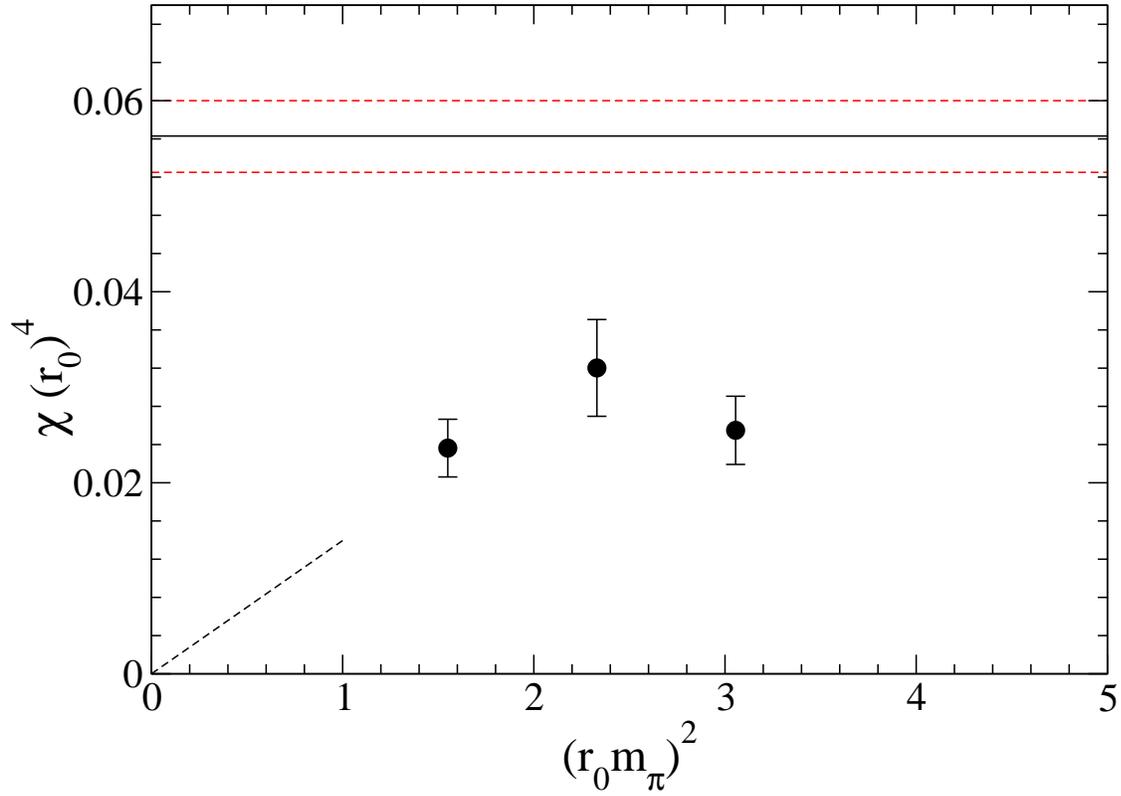}
\end{center}
\caption{Topological charge susceptibility for $N_f=2$ (filled circles) and quenched (solid line and horizontal dashed lines) simulations shown in Figure~\ref{fig:top charge}. The dashed line is the chiral perturbation theory prediction, Equation~\ref{eq:chiral pt chi}, with $r_0 f$ evaluated from~\cite{Aoki:2004ht}. Results are given in terms of the Sommer scale, $r_0$, for convenience.}
\label{fig:chi}
\end{figure}
\clearpage
\newpage

\begin{figure}
\begin{center}
\includegraphics[width=\textwidth]{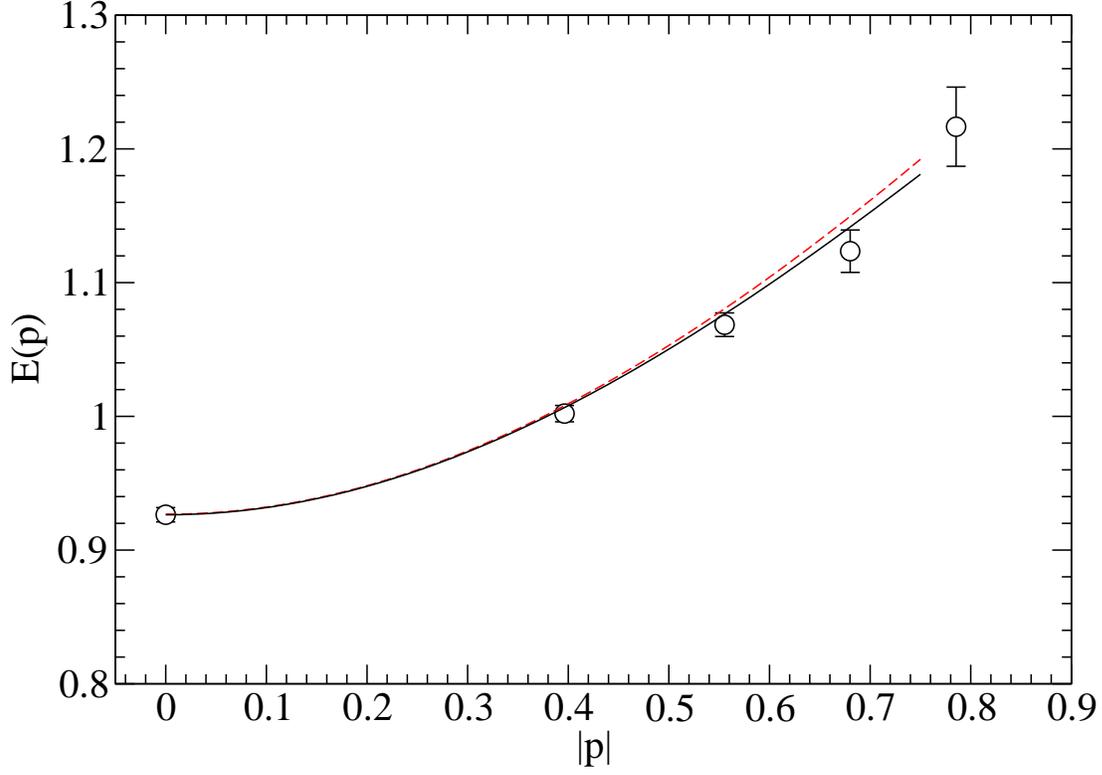}
\end{center}
\caption{The nucleon energy from a fit to Equation~\ref{eq:even 2pt}. $|p| = \frac{2\pi}{L} \sqrt{n_x^2 + n_y^2 + n_z^2}$ where $n_i = 0, \pm1,\pm 2$. The dashed line is plotted from the continuum formula, $E(p) = \sqrt{\vec p^2 + m_N^2}$, and the solid one is the same except $p_i$ is replaced by $\sin{(p_i)}$. All quantities are shown in lattice units. $m_{sea}=0.04$, $a^{-1}\approx 1.7$ GeV.}
\label{fig:energy}
\end{figure}
\clearpage
\newpage

\begin{figure}
\begin{center}
\includegraphics[width=\textwidth]{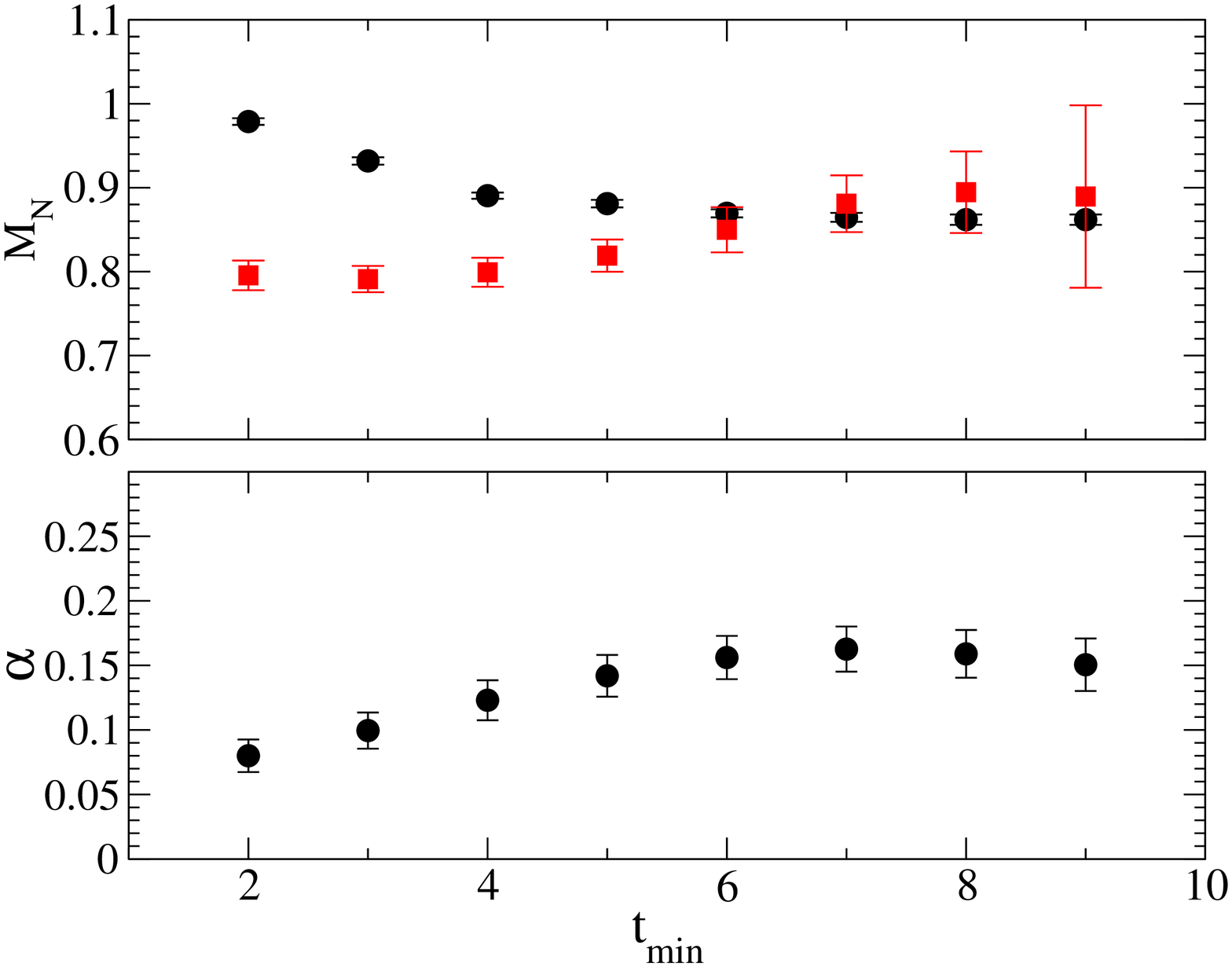}
\end{center}
\caption{Upper panel: nucleon mass from single particle fits to the CP-even (cirlcles) and odd (squares) parts of the two-point correlation function; $t_{min}$ is the value of the smallest time slice used in the fit. Lower panel: mixing angle $\alpha$ from the simple ratio of the same CP-odd and even parts of the two-point correlation function. $m_{sea}=0.03$.}
\label{fig:alpha m=0.03}
\end{figure}
\clearpage
\newpage

\begin{figure}
\begin{center}
\includegraphics[width=\textwidth]{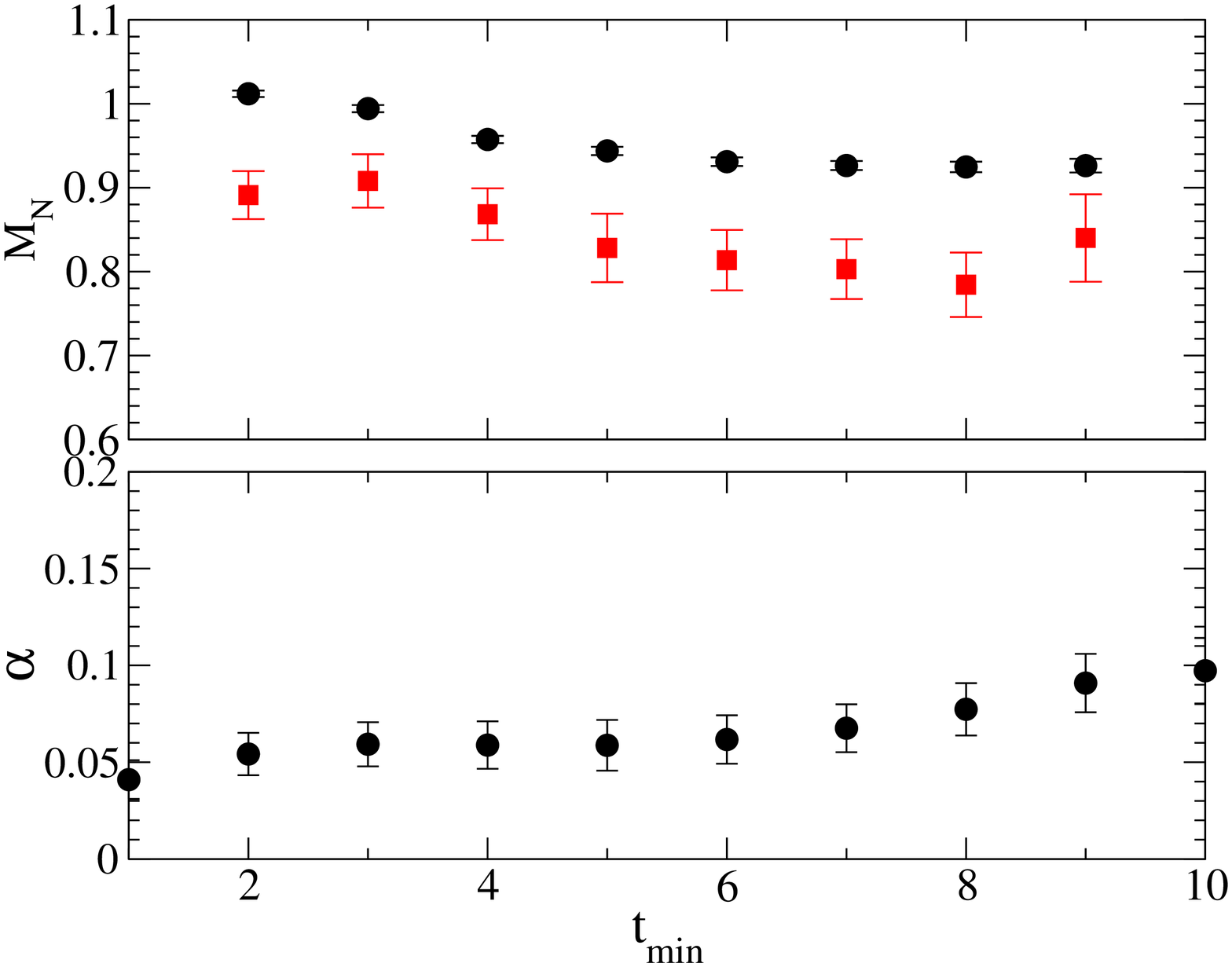}
\end{center}
\caption{Upper panel: nucleon mass from single particle fits to the CP-even (cirlcles) and odd (squares) parts of the two-point correlation function; $t_{min}$ is the value of the smallest time slice used in the fit. Lower panel: mixing angle $\alpha$ from the simple ratio of the same CP-odd and even parts of the two-point correlation function. $m_{sea}=0.04$.}
\label{fig:alpha m=0.04}
\end{figure}
\clearpage
\newpage

\begin{figure}
\begin{center}
\includegraphics[width=\textwidth]{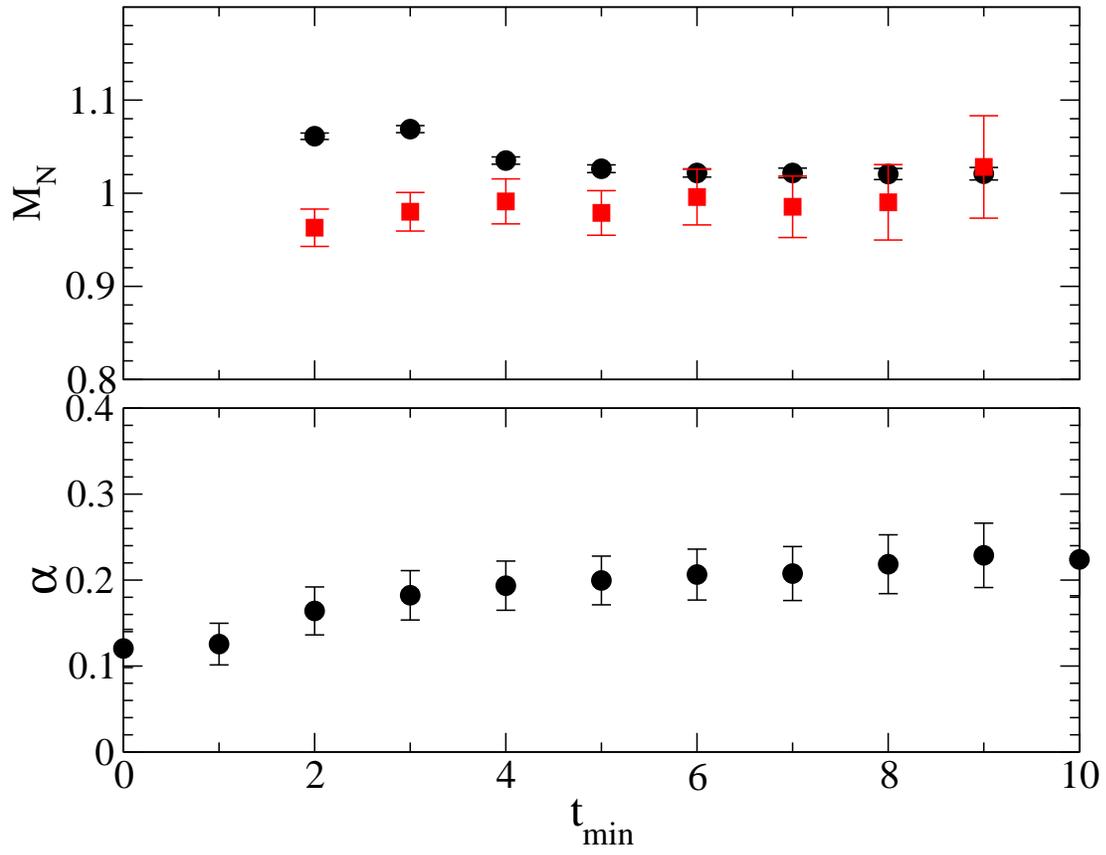}
\end{center}
\caption{ Same as Figure~\ref{fig:alpha m=0.04}, but for the quenched simulation described in the text.}
\label{fig:alpha quenched}
\end{figure}
\clearpage

\begin{figure}
\begin{center}
\includegraphics[width=\textwidth]{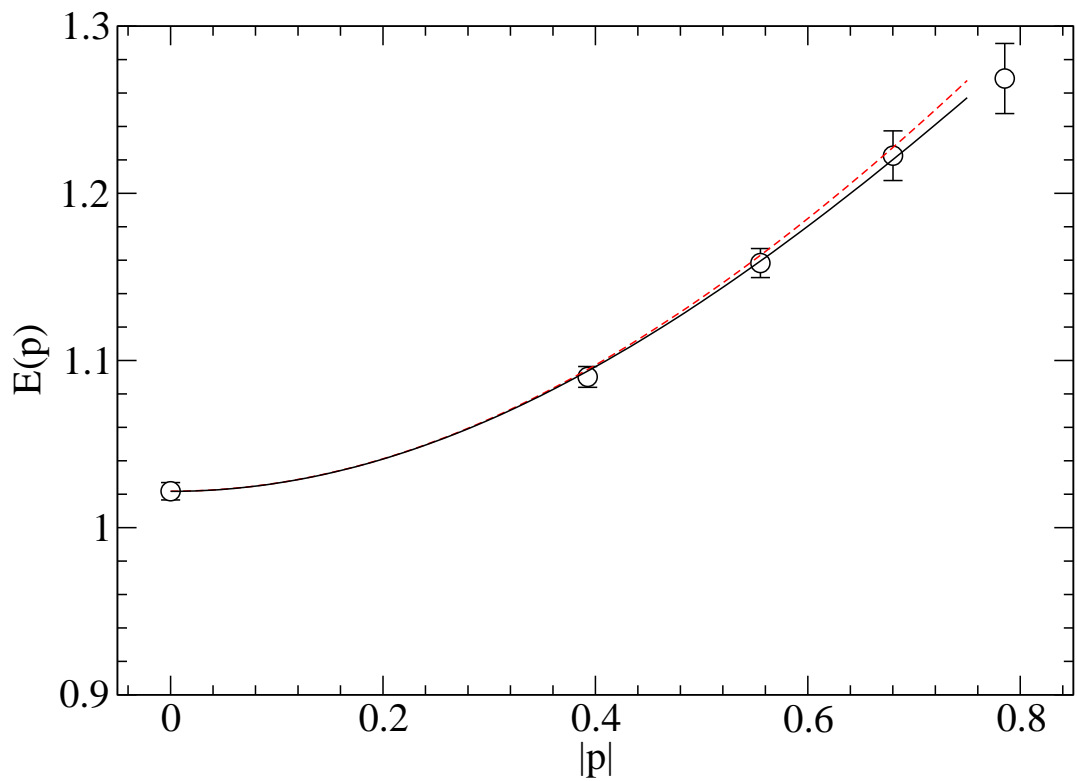}
\end{center}
\caption{Same as Figure~\ref{fig:energy}, except for the quenched ensemble described in the text.}
\label{fig:energy quenched}
\end{figure}
\clearpage
\newpage

\newpage
\begin{figure}
\begin{center}
\includegraphics[width=\textwidth]{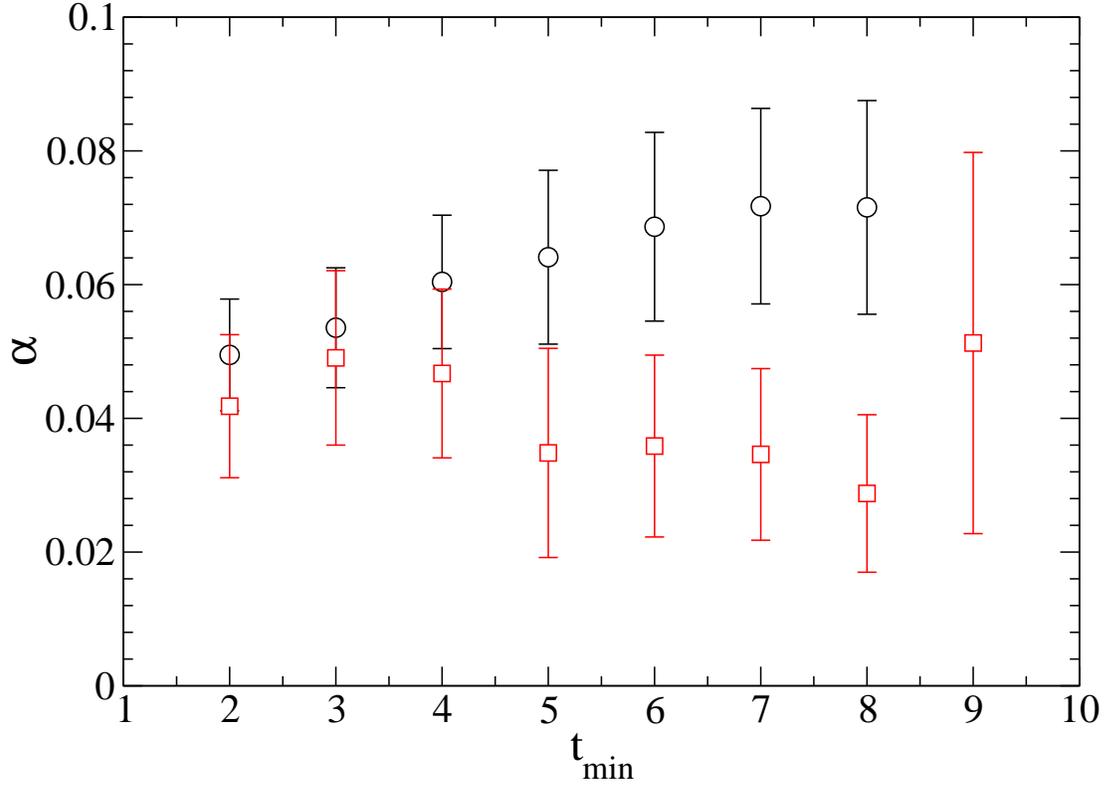}
\end{center}
\caption{The mixing angle $\alpha$ computed from fits of the CP even and odd parts of the nucleon two-point function to Equations~\ref{eq:two-point even} and ~\ref{eq:two-point odd}. Results are shown for $m_N$ fixed in the latter case to the value from the $\theta=0$ part (squares) and for $m_N$ a free parameter. $m_{sea}=0.04$.}
\label{fig:alpha-fixed}
\end{figure}
\clearpage
\newpage

\begin{figure}
\begin{center}
\includegraphics[width=\textwidth]{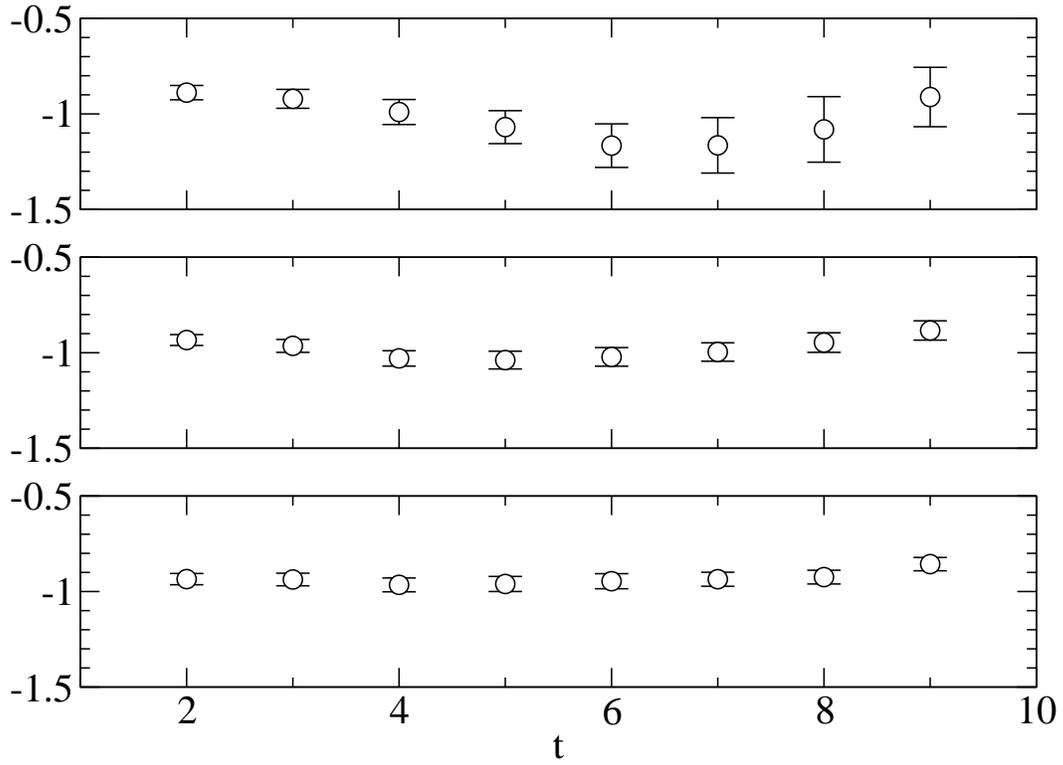}
\end{center}
\caption{Ratio of three-point functions given in Equation~\ref{eq:magnetic ratio} that yields the magnetic dipole moment of the neutron in the limit $q^2\to0$. Plots are shown for units of lattice momenta, $\vec q=(1,0,0)$, $(1,1,0)$, and $(1,1,1)$, and permutations, in the lower, middle, and upper panels respectively. $m_{sea}=0.03$.}
\label{fig:magnetic ratio}
\end{figure}
\clearpage
\newpage

\begin{figure}
\begin{center}
\includegraphics[width=\textwidth]{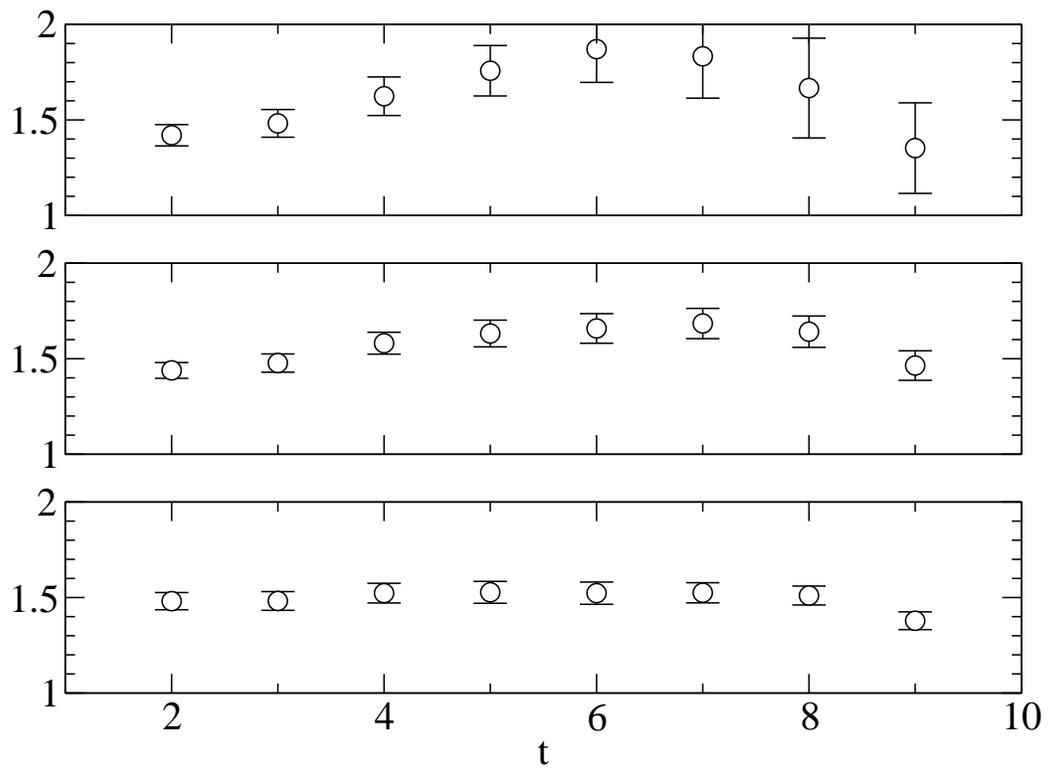}
\end{center}
\caption{Same as in Figure~\ref{fig:magnetic ratio} but for the proton.}
\end{figure}
\clearpage
\newpage

\begin{figure}
\begin{center}
\includegraphics[width=\textwidth]{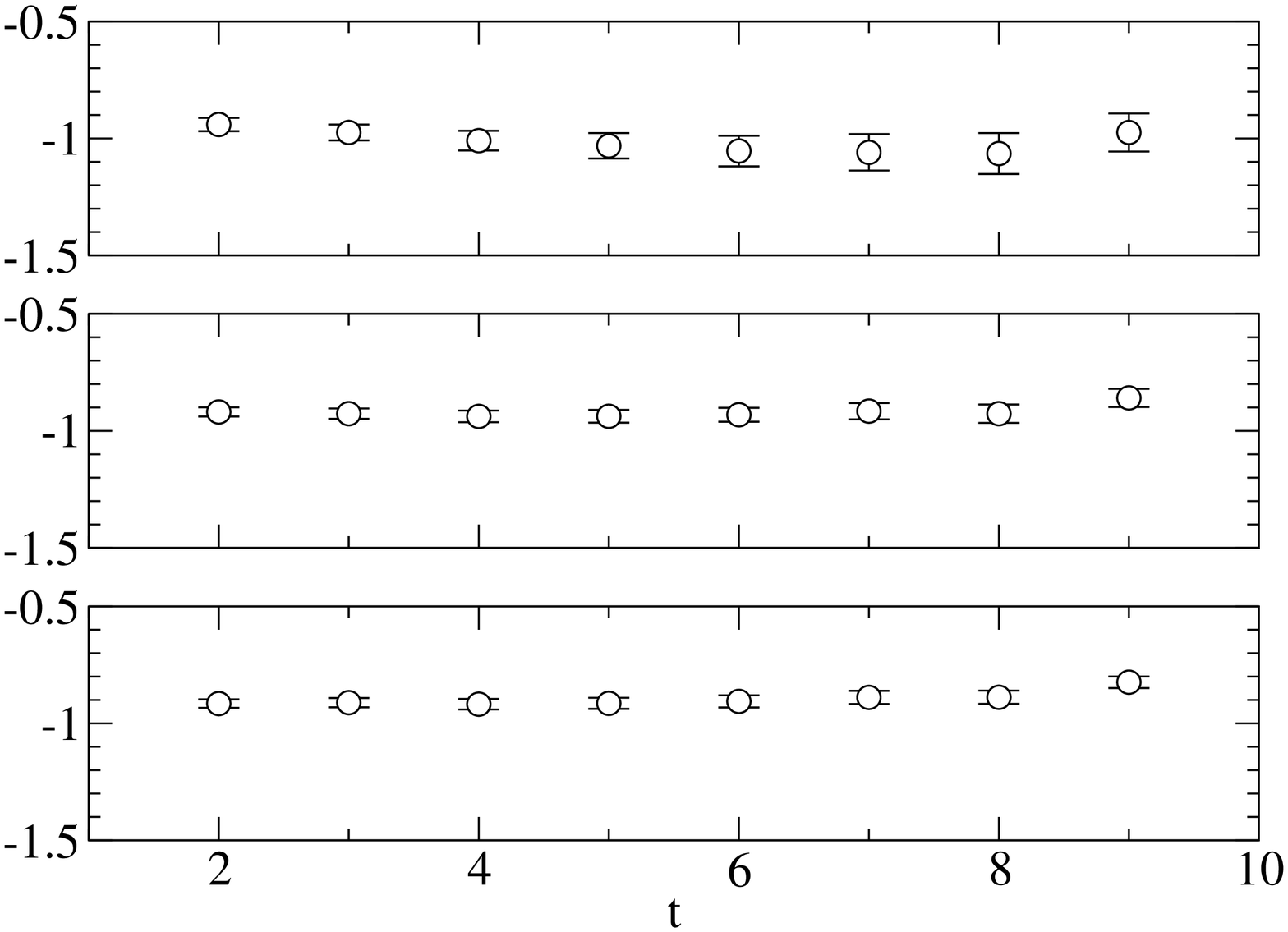}
\end{center}
\caption{Same as in Figure~\ref{fig:magnetic ratio} but $m_{sea}=0.04$.}
\label{fig:magnetic ratio m04}
\end{figure}
\clearpage
\newpage

\begin{figure}
\begin{center}
\includegraphics[width=\textwidth]{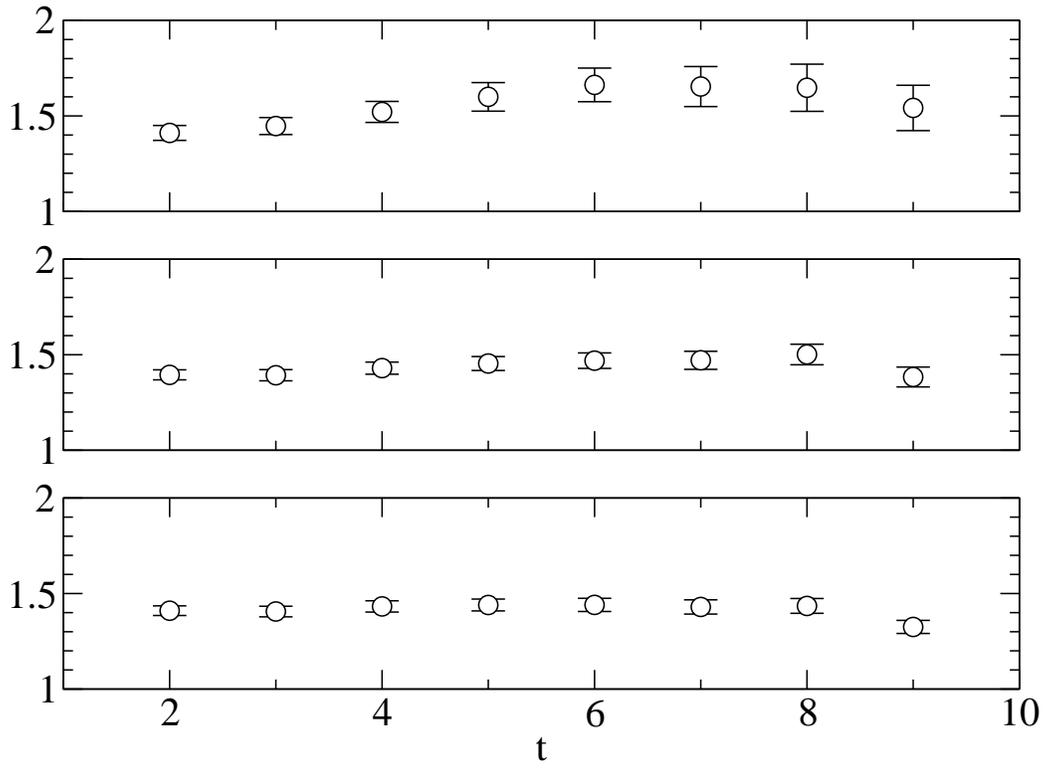}
\end{center}
\caption{Same as in Figure~\ref{fig:magnetic ratio} but for the proton, and $m_{sea}=0.04$.}
\end{figure}
\clearpage
\newpage

\begin{figure}
\begin{center}
\includegraphics[width=\textwidth]{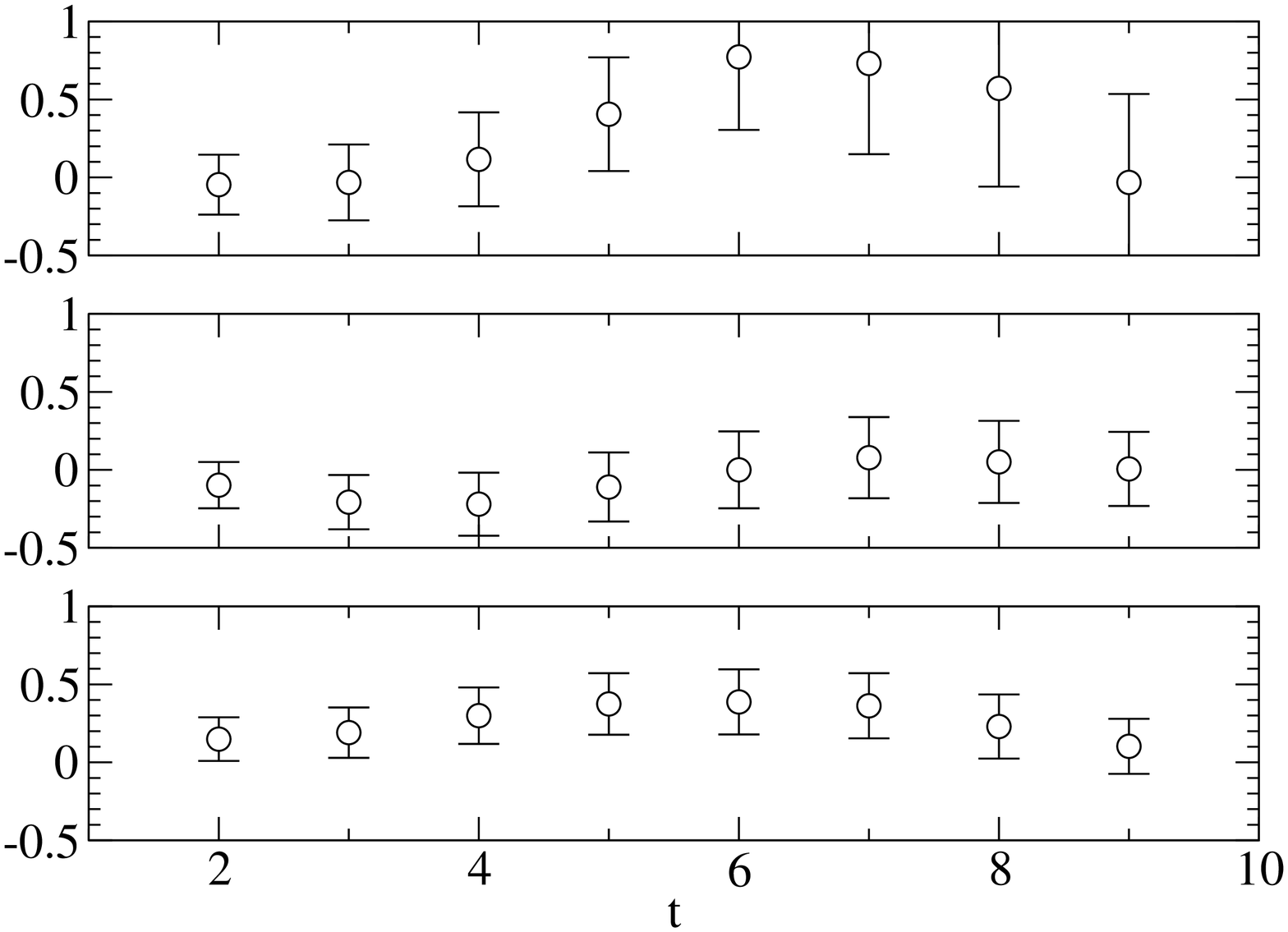}
\end{center}
\caption{Same as in Figure~\ref{fig:magnetic ratio}, but for the electric dipole moment of the neutron (Equation~\ref{eq:f2-f3 ratio}). The mixing with the $F_1$ and $F_2$ terms has not been subtracted. $m_{sea}=0.03$.}
\label{fig:f2-f3 ratio m03}
\end{figure}
\clearpage
\newpage
\begin{figure}
\begin{center}
\includegraphics[width=\textwidth]{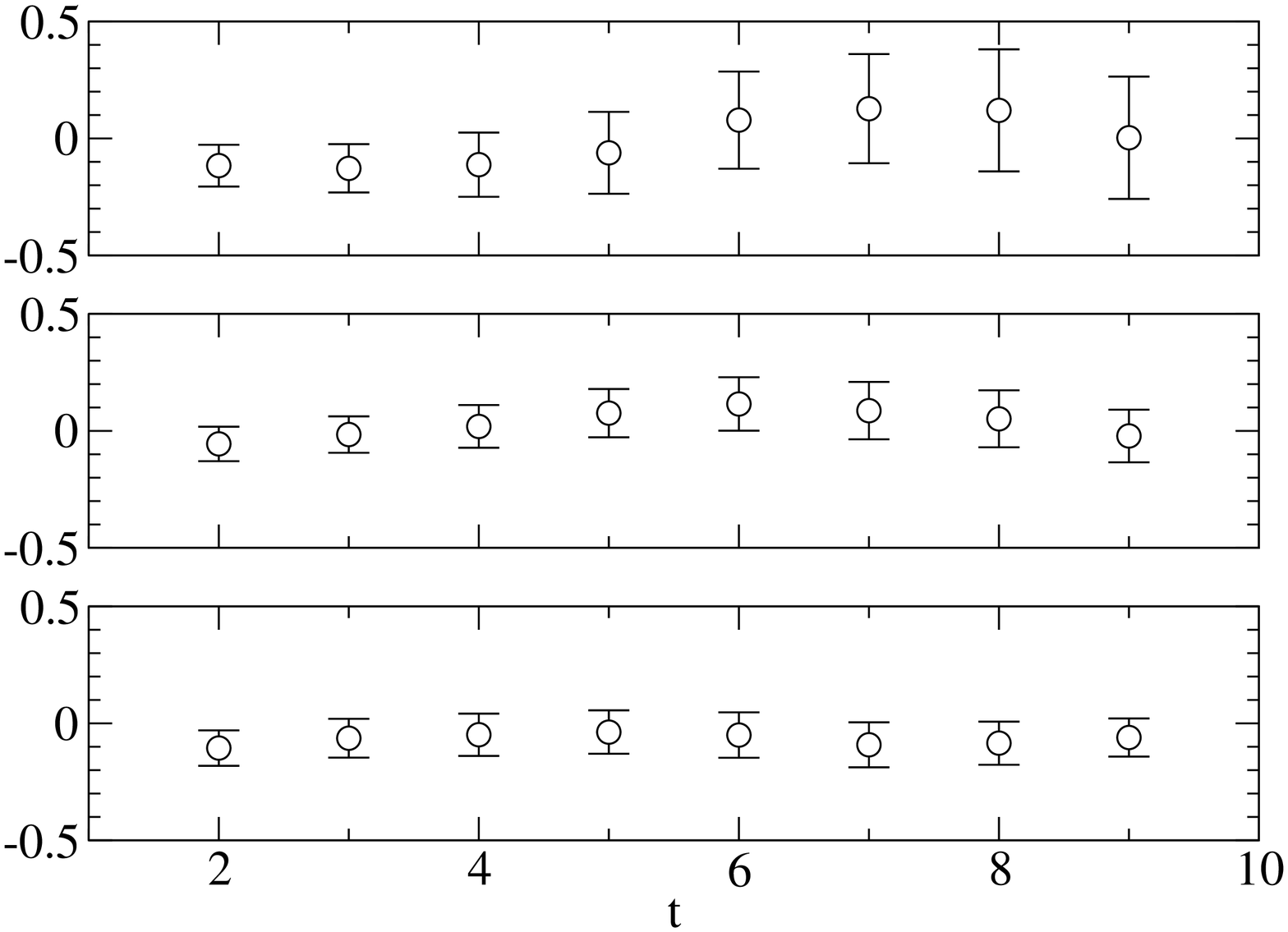}
\end{center}
\caption{Same as in Figure~\ref{fig:magnetic ratio}, but for the electric dipole moment of the neutron (Equation~\ref{eq:f2-f3 ratio}). The mixing with the $F_1$ and $F_2$ terms has not been subtracted. $m_{sea}=0.04$.}
\label{fig:f2-f3 ratio}
\end{figure}
\clearpage
\newpage

\begin{figure}
\begin{center}
\includegraphics[width=\textwidth]{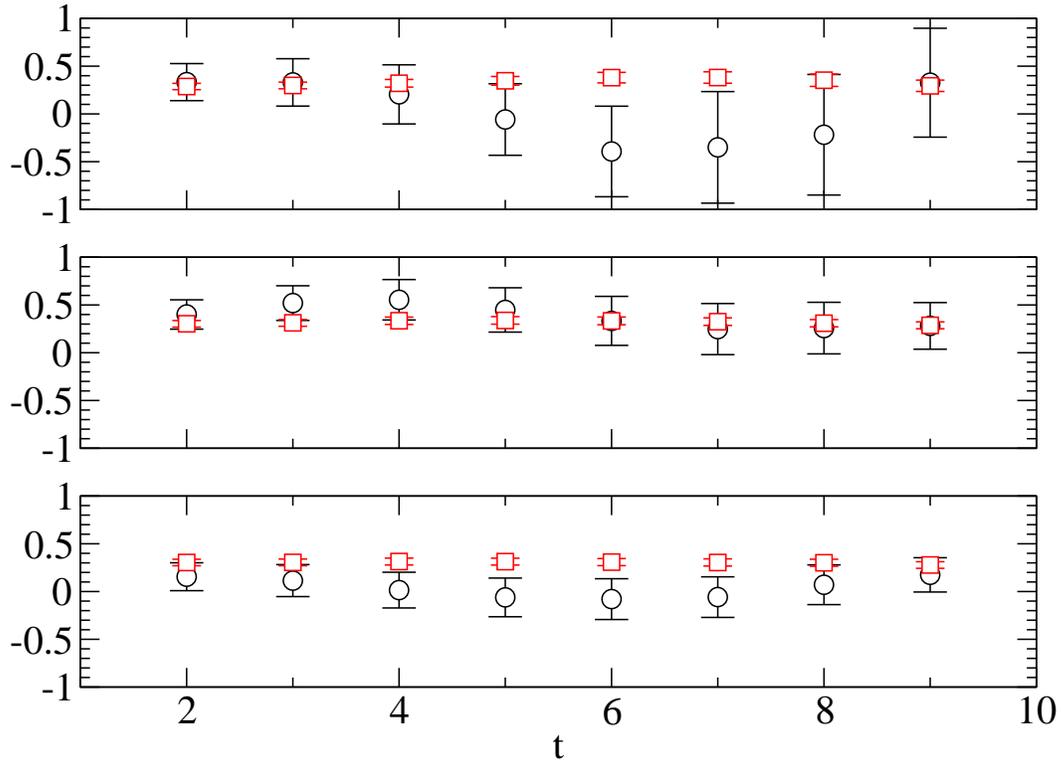}
\end{center}
\caption{The ratio $(F_3(q^2)/2m)/G_E(q^2)$ (Equation~\ref{eq:electric dipole moment form factor}). $q^2$ increases from bottom to top panel. In the limit $q^2\to0$ this ratio yields the electric dipole moment of the neutron. Also shown is the subtraction term (squares) in Equation~\ref{eq:electric dipole moment form factor}. $m_{sea}=0.03$}
\label{fig:edm ratio}
\end{figure}
\clearpage
\newpage
\begin{figure}
\begin{center}
\includegraphics[width=\textwidth]{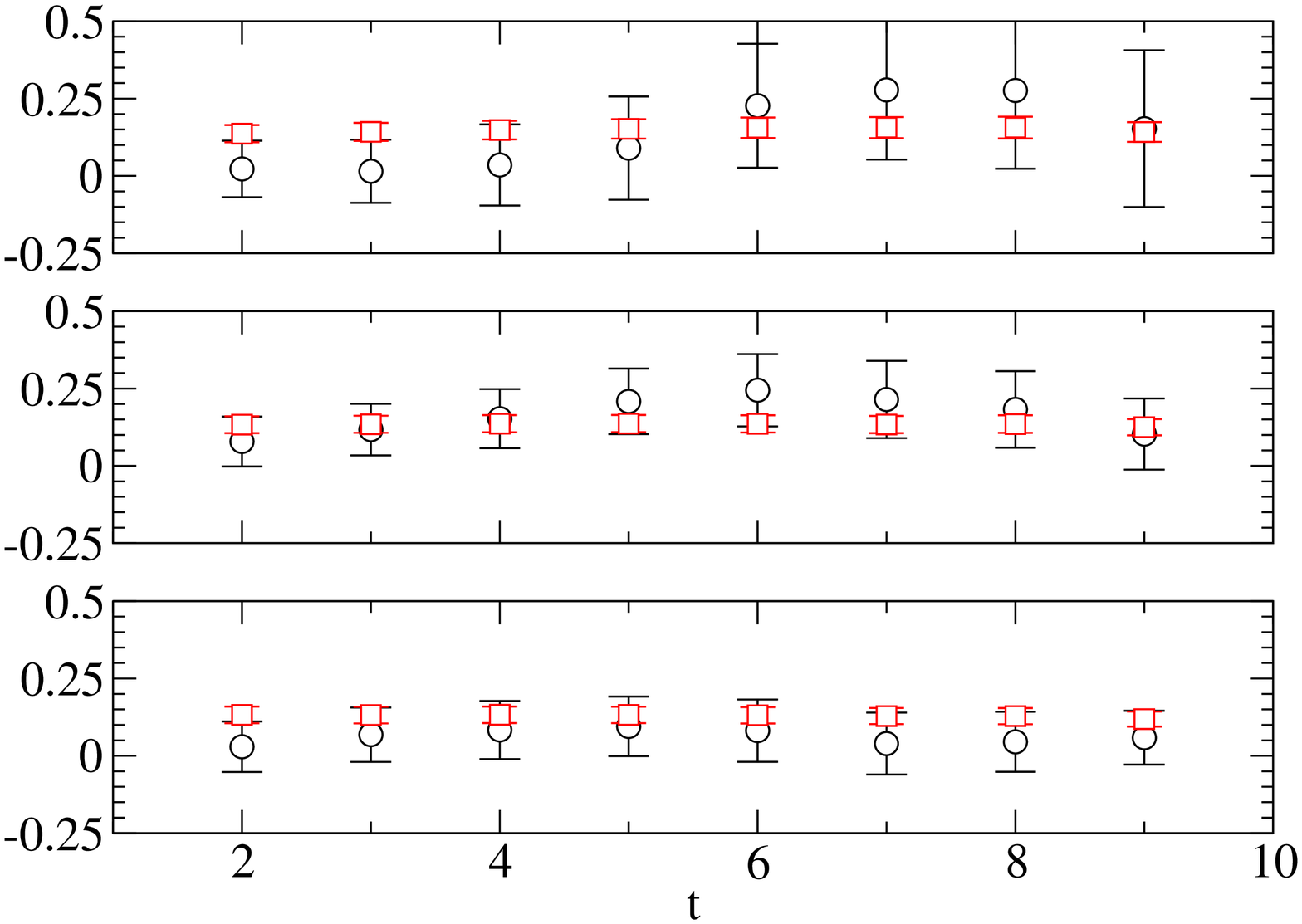}
\end{center}
\caption{Same as in Figure~\ref{fig:edm ratio}, but $m_{sea}=0.04$.}
\label{fig:edm ratio 04}
\end{figure}
\clearpage
\newpage

\begin{figure}
\begin{center}
\includegraphics[width=\textwidth]{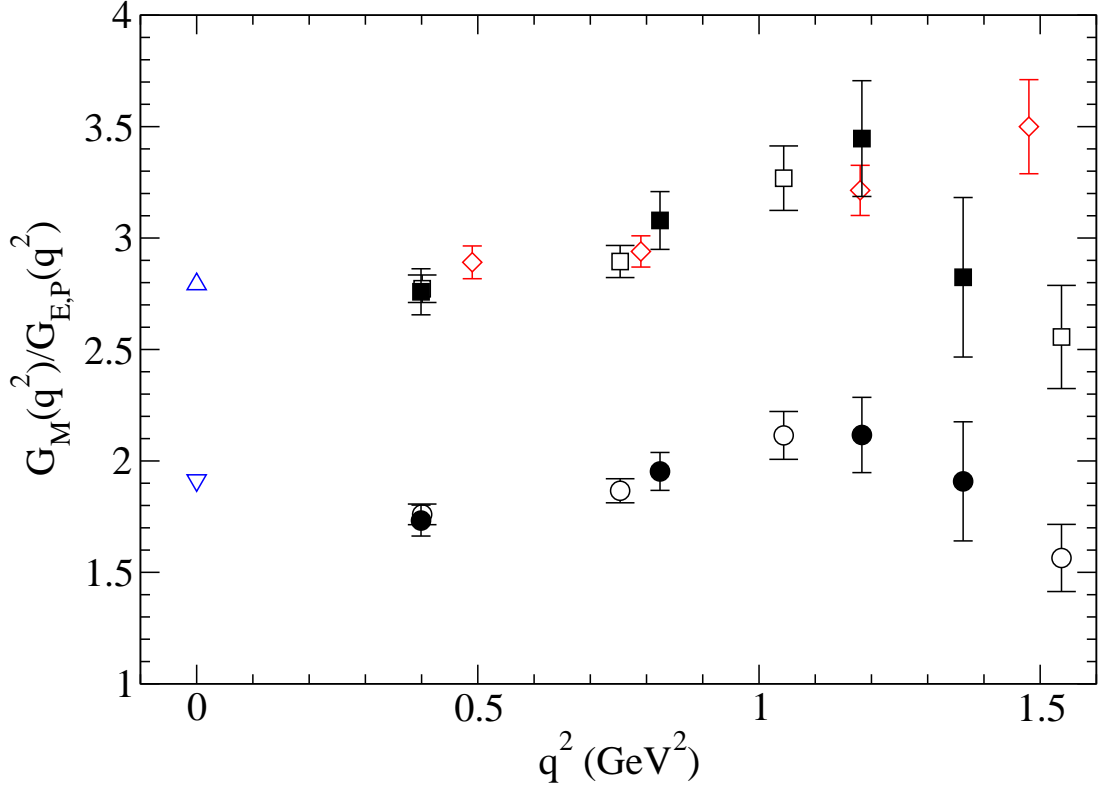}
\end{center}
\caption{Ratios of the proton (squares) and neutron (circles) magnetic form factors to the proton electric form factor (Equation~\ref{eq:gm/ge} times $E+m$). The limit $q^2\to0$ yields the magnetic moments. The absolute value is plotted for the neutron for comparison. $m_{sea}=0.03$ (filled symbols) and $0.04$ (open symbols). The diamonds are experimental data points~\cite{Jones:1999rz} where we have added quoted statistical and systematic errors in quadrature. The triangles at $q^2=0$ are the experimentally measured magnetic moments \cite{Eidelman:2004wy} (the absolute value for the neutron (lower triangle) is shown).}
\label{fig:gm/ge}
\end{figure}
\clearpage
\newpage


\fi

\end{document}